\documentclass{sigchi}

\CopyrightYear{2020}
\setcopyright{acmlicensed}
\doi{https://doi.org/10.1145/3313831.XXXXXXX}
\isbn{978-1-4503-6708-0/20/04}
\conferenceinfo{CHI'20,}{April  25--30, 2020, Honolulu, HI, USA}
\acmPrice{\$15.00}

\makeatletter
\def\@copyrightspace{\relax}
\makeatother

\usepackage{balance}       % to better equalize the last page
\usepackage{graphics}      % for EPS, load graphicx instead 
\usepackage[T1]{fontenc}   % for umlauts and other diaeresis
\usepackage{txfonts}
\usepackage{mathptmx}
\usepackage[pdflang={en-US},pdftex]{hyperref}
\usepackage{color}
\usepackage{booktabs}
\usepackage{textcomp}

\usepackage{microtype}        % Improved Tracking and Kerning
\usepackage{ccicons}          % Cite your images correctly!
\usepackage{todonotes}

\usepackage{subfig}
\usepackage[export]{adjustbox}
\graphicspath{{figures/}{pictures/}{images/}{./}{imgs/}} % where to search for the images

\def\plaintitle{Truncating the Y-Axis: Threat or Menace?}

\def\emptyauthor{}
\def\plainkeywords{Information visualization; Deceptive Visualization}

\makeatletter
\def\url@leostyle{%
  \@ifundefined{selectfont}{
    \def\UrlFont{\sf}
  }{
    \def\UrlFont{\small\bf\ttfamily}
  }}
\makeatother
\def\pprw{8.5in}
\def\pprh{11in}

\definecolor{linkColor}{RGB}{6,125,233}
\hypersetup{%
  pdftitle={\plaintitle},
  pdfauthor={\emptyauthor},
  pdfkeywords={\plainkeywords},
  pdfdisplaydoctitle=true, % For Accessibility
  bookmarksnumbered,
  pdfstartview={FitH},
  colorlinks,
  citecolor=black,
  filecolor=black,
  linkcolor=black,
  urlcolor=linkColor,
  breaklinks=true,
  hypertexnames=false
}

\begin{document}

\title{\plaintitle}

\emptyauthor{}
\numberofauthors{3}
\author{%
  \alignauthor{Michael Correll\\
    \affaddr{Tableau Research}\\
    \email{mcorrell@tableau.com}}\\
  \alignauthor{Enrico Bertini\\
    \affaddr{New York University}\\
    \email{enrico.bertini@nyu.edu}}\\
  \alignauthor{Steven Franconeri\\
    \affaddr{Northwestern University}\\
    \email{franconeri@northwestern.edu}}\\
}

\maketitle

\begin{abstract}
Bar charts with y-axes that don't begin at zero can visually exaggerate effect sizes. However, advice for whether or not to truncate the y-axis can be equivocal for other visualization types. In this paper we present examples of visualizations where this y-axis truncation can be beneficial as well as harmful, depending on the communicative and analytic intent. We also present the results of a series of crowd-sourced experiments in which we examine how y-axis truncation impacts subjective effect size across visualization types, and we explore alternative designs that more directly alert viewers to this truncation. We find that the subjective impact of axis truncation is persistent across visualizations designs, even for designs with explicit visual cues that indicate truncation has taken place. We suggest that designers consider the scale of the meaningful effect sizes and variation they intend to communicate, regardless of the visual encoding.
\end{abstract}

% ACM Classfication



% Author Keywords

% Print the classficiation codes

%% \section{Introduction} %for journal use above \firstsection{..} instead

\section{Introduction}
\begin{figure*}
    \centering
    \subfloat[]{
        \includegraphics[width=0.6\columnwidth]{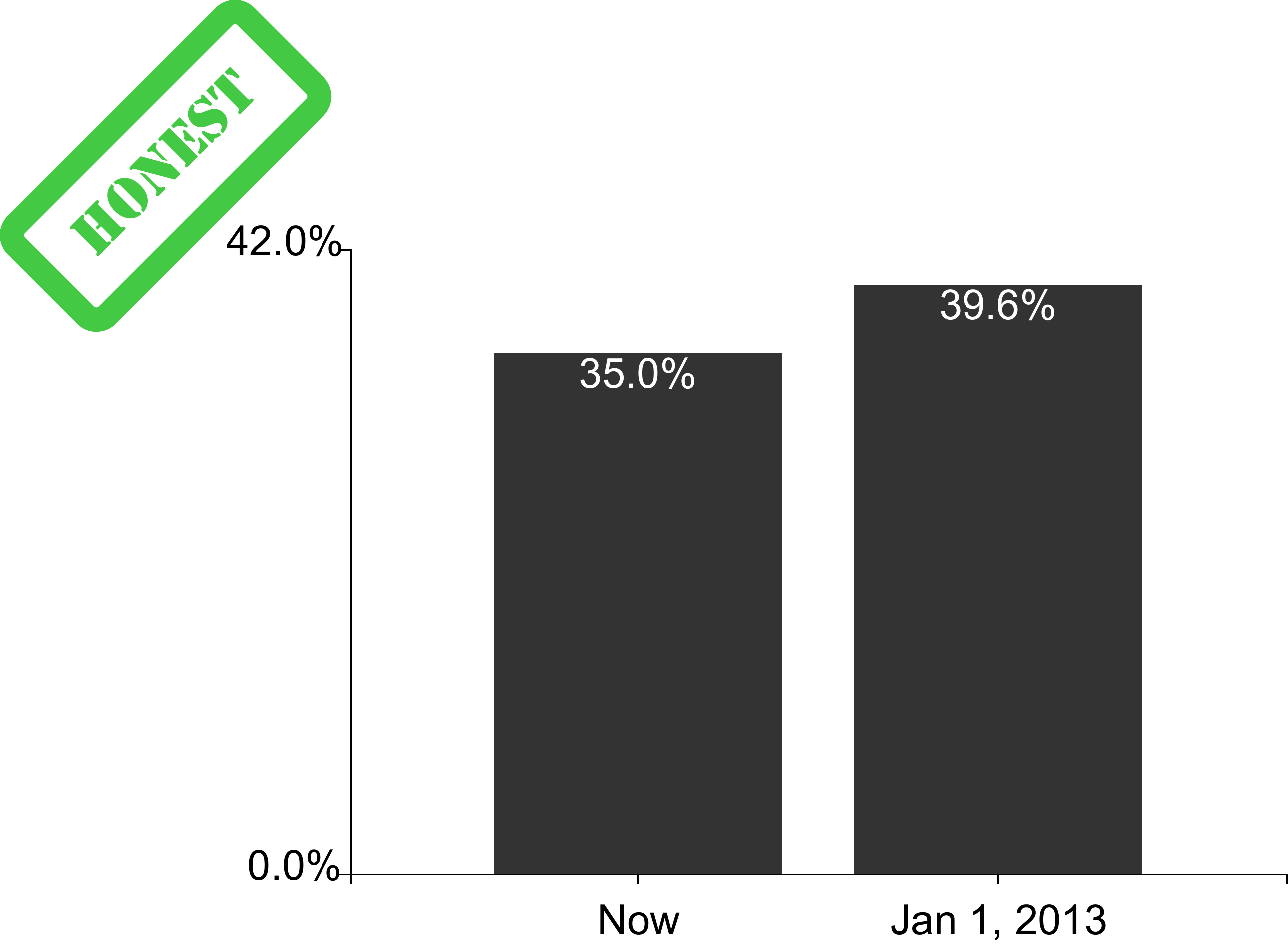}
    }
    ~
    \subfloat[]{
        \includegraphics[width=0.6\columnwidth]{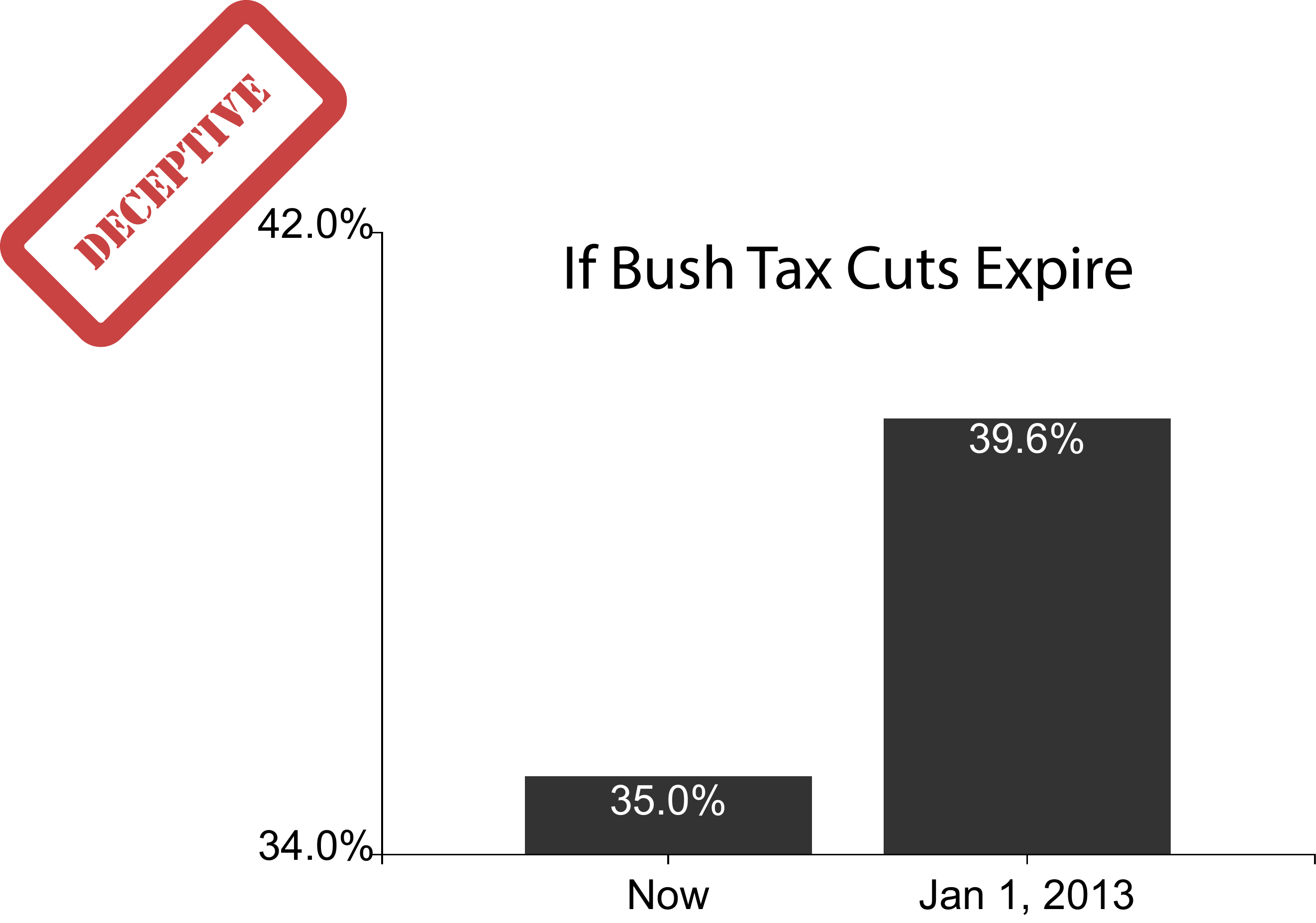}
        \label{fig:fox}
    }
    
    \subfloat[]{
        \includegraphics[width=0.6\columnwidth]{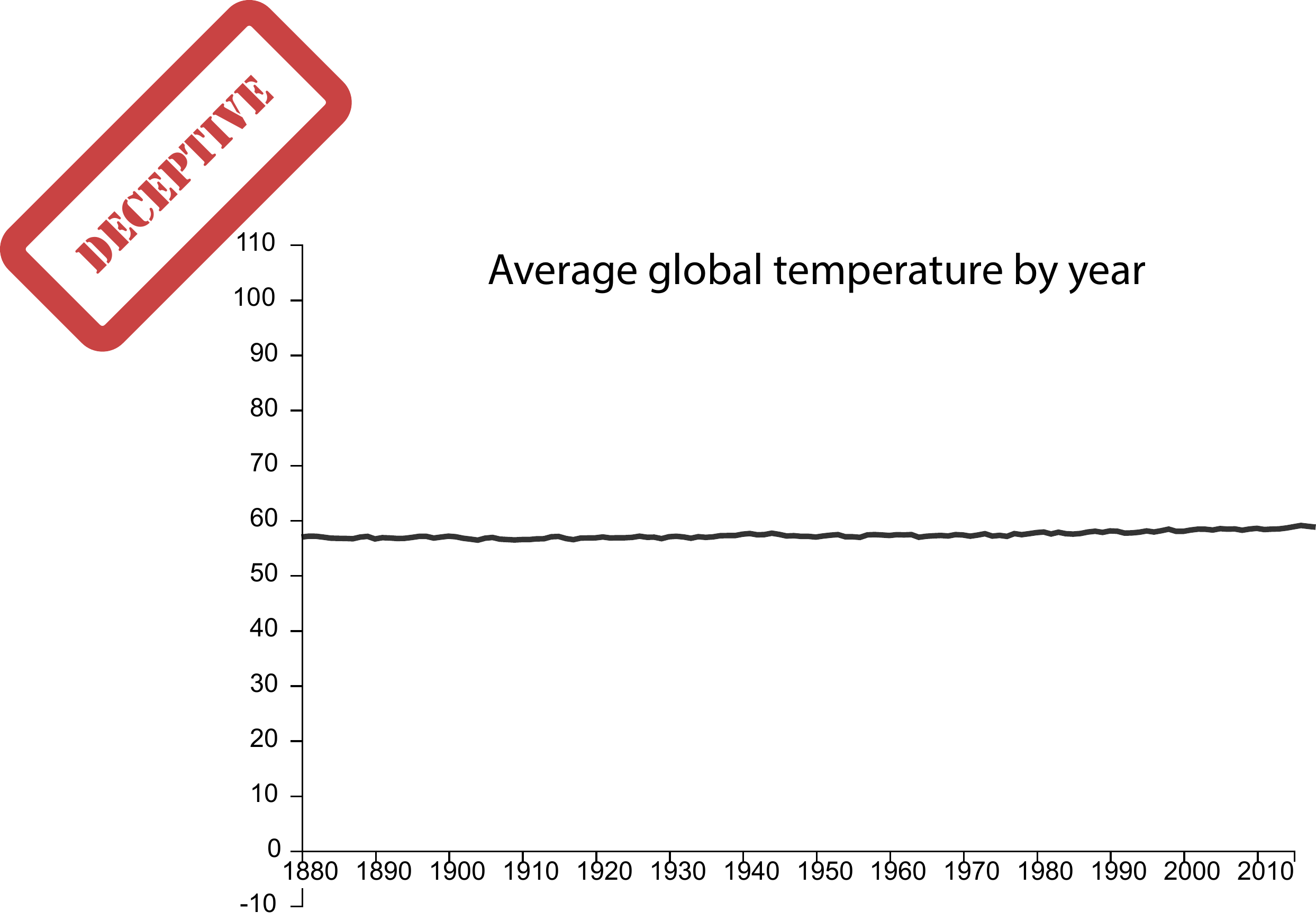}
        \label{fig:nro}
    }
    ~
    \subfloat[]{
        \includegraphics[width=0.6\columnwidth]{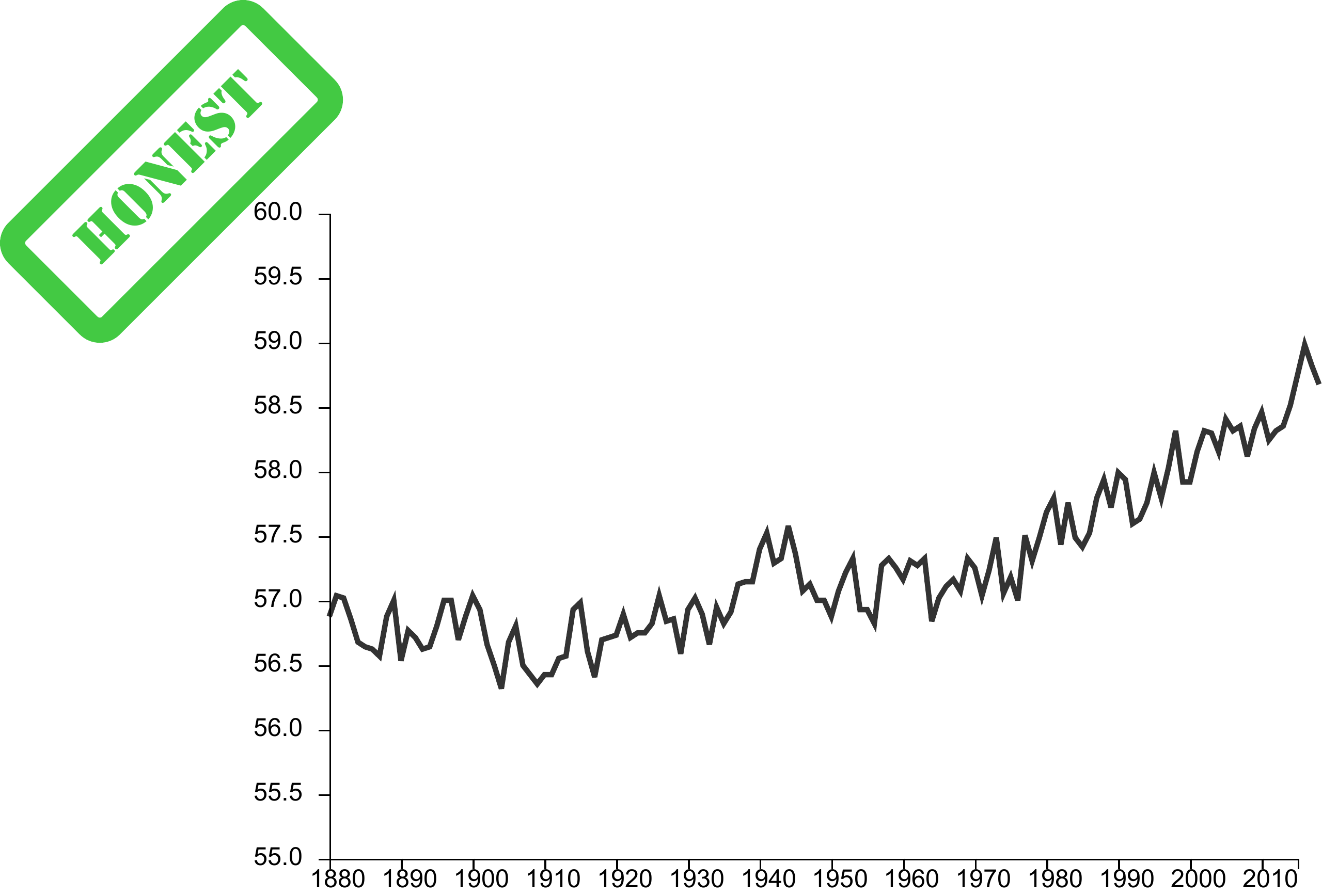}
    }
    \caption{Two charts that manipulate the y-axis in ways that seem deceptive. In Fig.~\protect\ref{fig:fox}, a reproduction of a chart aired on Fox News\protect\cite{foxbad}, the second bar is 6 times taller than the first bar, even though there is only a 4.6\% increase in tax rate (ratio of 1.13 to 1). In Fig.~\protect\ref{fig:nro}, a reproduction of a chart tweeted by the \emph{National Review} magazine\protect\cite{nrotweet}, taken from the Power Line blog, a warming climate is obscured by making the y-axis start at 0 degrees Fahrenheit, compressing the trend into illegibility. }
    \label{fig:badcharts}
\end{figure*}
    
Satirical headlines with ``threat or menace'' are used to satirize topics about which the people appear to have already made up their minds~\cite{threat}. Starting the quantitative axis of a bar chart from a value other than zero appears to be one such anathema, and is considered one of the cardinal sins of information visualization. By starting the axis from a value other than zero, the designer \emph{truncates} the range of y-values, over-emphasizing minute differences between values that would otherwise appear very similar in a zero-baseline chart. Bar charts truncated in this manner have been called ``biased''~\cite{szafir2018good}, ``dishonest,''~\cite{foxbad}, ``deceptive''~\cite{o2018testing,pandey2015deceptive}, ``lying with statistics''~\cite{huff1993lie}, and ``the worst of crimes in data visualisation''~\cite{economist}, with this exaggeration quantified in Tufte's ``lie factor''~\cite{tufte}. Prior work has shown that truncation and exaggeration in axes results in quantifiable differences in how people interpret the size and significance of effects~\cite{o2018testing,pandey2015deceptive,witt2019graph}, affects judgments of correlation~\cite{cleveland1982variables}, and makes trends appear more subjectively ``threatening''~\cite{berger2005slippery}. A prescription against non-zero baselines for bar charts is encoded as a hard constraint in automated visualization design tools like Draco~\cite{moritz2019formalizing}.

In many guidelines concerning y-axes, bar charts are specifically mentioned as being vulnerable to truncation. By contrast, the injunction against truncating the y-axis is often considered less pressing for line charts (compare the examples in Fig.~\ref{fig:badcharts}). Bar charts use length to encode value, and are often used to afford the quick comparison of individual values. Truncating the y-axis of a bar chart breaks the visual convention that the difference in the height of the bars is proportional to the difference in values, and so is misleading from an encoding standpoint~\cite{correll2017black}. By contrast, when trends, rather than individual values, are important components of the intended messages, there are some cases where \emph{not} truncating the y-axis is perceived as deceptive (Fig.~\ref{fig:nro}).

Despite this negativity, there is relatively little empirical work on how y-axis truncation inflates judgments across different visual encodings~\cite{pandey2015deceptive,ritchie2019,witt2019graph}. In this paper, we summarize the current debate over y-axis truncation. We also present the results of a crowd-sourced experiment investigating the impact of y-axis truncation on subjective assessments of data across different visualization designs. We find that the ability of y-axis truncation to exaggerate effect sizes is not limited to typical bar charts, but extends to designs proposed specifically for indicating that truncation has taken place. Our results suggest that the designer therefore has a great deal of control over the perceived effect size in data. There is therefore not a clear binary distinction between ``deceptive'' versus ``truthful'' y-axis presentations: designers must take into account the range and magnitude of effect sizes they wish to communicate at a per-data and per-task level.

\section{Existing Guidelines About Y-Axis Truncation}

\begin{figure}
    \centering
    \includegraphics[width=0.95\columnwidth]{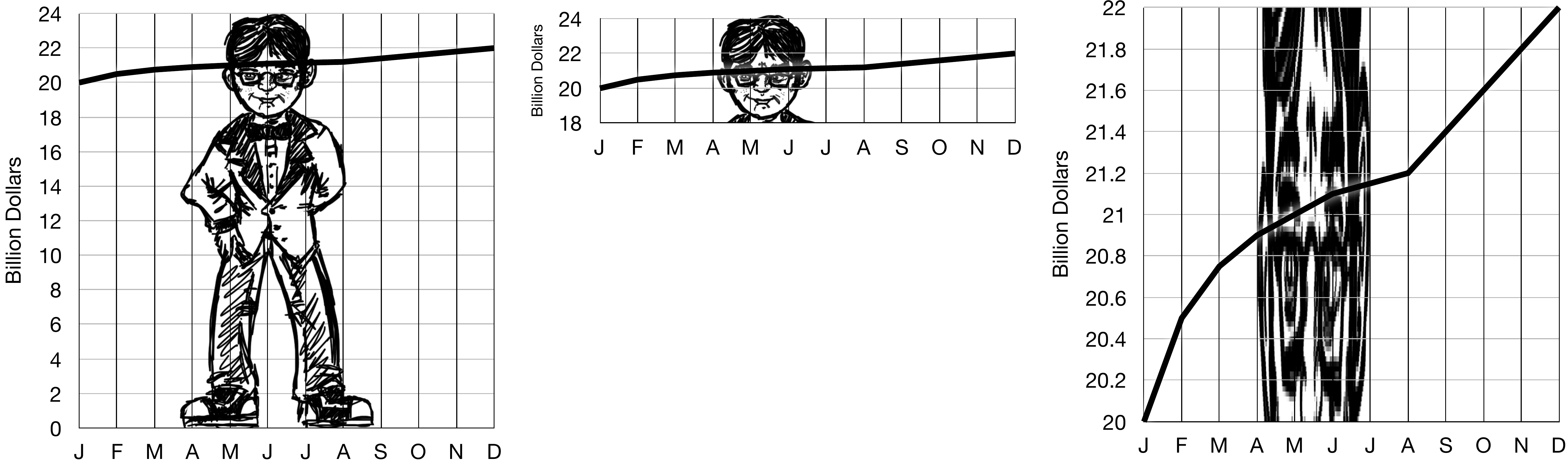}
    \caption{Recreation of a visual argument against ``Gee Whiz Charts,'' from Huff~\protect\cite{huff1993lie}. For two charts of identical size, starting the y-axis at a point other than $0$ creates two sorts of deceptions, for Huff: chopping off most of the context of the chart, and then exaggerating the remaining context. Note that this argument applies broadly to most charts with quantitative y-axes, not just line charts.}
    \label{fig:howtolie}
\end{figure}
When it is permissible to truncate the y-axis is a subject of continuous and active debate. Much of this debate occurs in the pages of books, or in informal channels like Twitter and blog posts, rather than in academic articles. In this section we summarize major positions in this debate, with the intent of synthesizing the major rationales behind existing guidelines.

Huff's~\cite{huff1993lie} \emph{How to Lie With Statistics} calls out charts with non-zero axes as ``Gee Whiz Graphs.'' After y-axis truncation:
\begin{quote}
    The figures are the same and so is the curve. It is the same graph. Nothing has been falsified-- except the impression that it gives. But what the hasty reader sees now is a national-income line that has climbed halfway up the paper in twelve months, all because most of the chart isn't there any more... a small rise has become, visually, a big one.
\end{quote}

Huff's advice is that therefore all charts of positive values should begin at 0, lest the designer deceive by making a trend appear more ``impressive'' than it ought to be (see Fig.~\ref{fig:howtolie}).

Brinton's~\cite{brinton1939graphic} chapter on ``Standards for Time Series Charts'' in \emph{Graphic Presentation} similarly claims:
\begin{quote}
    The amount scale should normally include the zero value or other principle point of reference. Departure from this rule should never be made except where there is a special reason for so doing.
\end{quote}

Although he includes an exception, and suggests using visual indicators (such as a ``torn paper'' metaphor~\cite{tornbaseline}) to indicate when an axis has been adjusted:
\begin{quote}
    When the interest of the reader is in the absolute amount of change rather than in the relative amount of change, it may be safe to omit the principal point of reference and the accompanying horizontal line[... w]hen the zero value or other principal point of reference is omitted the fact should be clearly indicated in a manner that will attract notice.
\end{quote}

More recent discussion on the issue has been less dogmatic, and focuses on how different graphs encode data in different ways, and for different purposes. Alberto Cairo, in \emph{How Charts Lie}~\cite{cairo2019}, proposes the following rule:
\begin{quote}
I usually advise a baseline of zero when the method of encoding is height or length. If the method of encoding is different, a zero baseline may not always be necessary. The encodings in a line chart are position and angle, and these don't get distorted if we set a baseline that is closer to the first data point.
\end{quote}

Similarly, Carl Bergstrom and Jevin West in their critical thinking website ``Calling Bullshit''~\cite{callingbullshit} hold that their principal of ``proportional ink'' (somewhat analogous to Tufte's ``lie factor''~\cite{tufte}) does not apply for line charts:
\begin{quote}
    [...]unlike bar charts, line graphs need not include zero on the dependent variable axis. Why not? The answer is that line charts don't use shaded volumes to indicate quantities; rather, they use positions that indicate quantities. The principle of proportional ink therefore does not apply, because the amount of ink is not used to indicate the magnitude of a variable. Instead, a line chart should be scaled so as to make the position of each point maximally informative, usually by allowing the axis to span the region not much larger than the range of the data values.
\end{quote}

Tufte himself, in a posting on his website~\cite{tufte}, seems to take a similar view, but narrows his exception to time series data rather than line charts in general:
\begin{quote}
    In general, in a time-series, use a baseline that shows the \emph{data}, not the zero point. If the zero point reasonably occurs in plotting the data, fine. But don't spend a lot of empty vertical space trying to reach down to the zero point at the cost of hiding what is going on in the data line itself.
\end{quote}

\begin{figure}[h!]
\centering
\subfloat[Statistical process charts rely on comparison to an expected value, and so deviations from that value, not from zero, are important]{
\includegraphics[width=0.7\columnwidth]{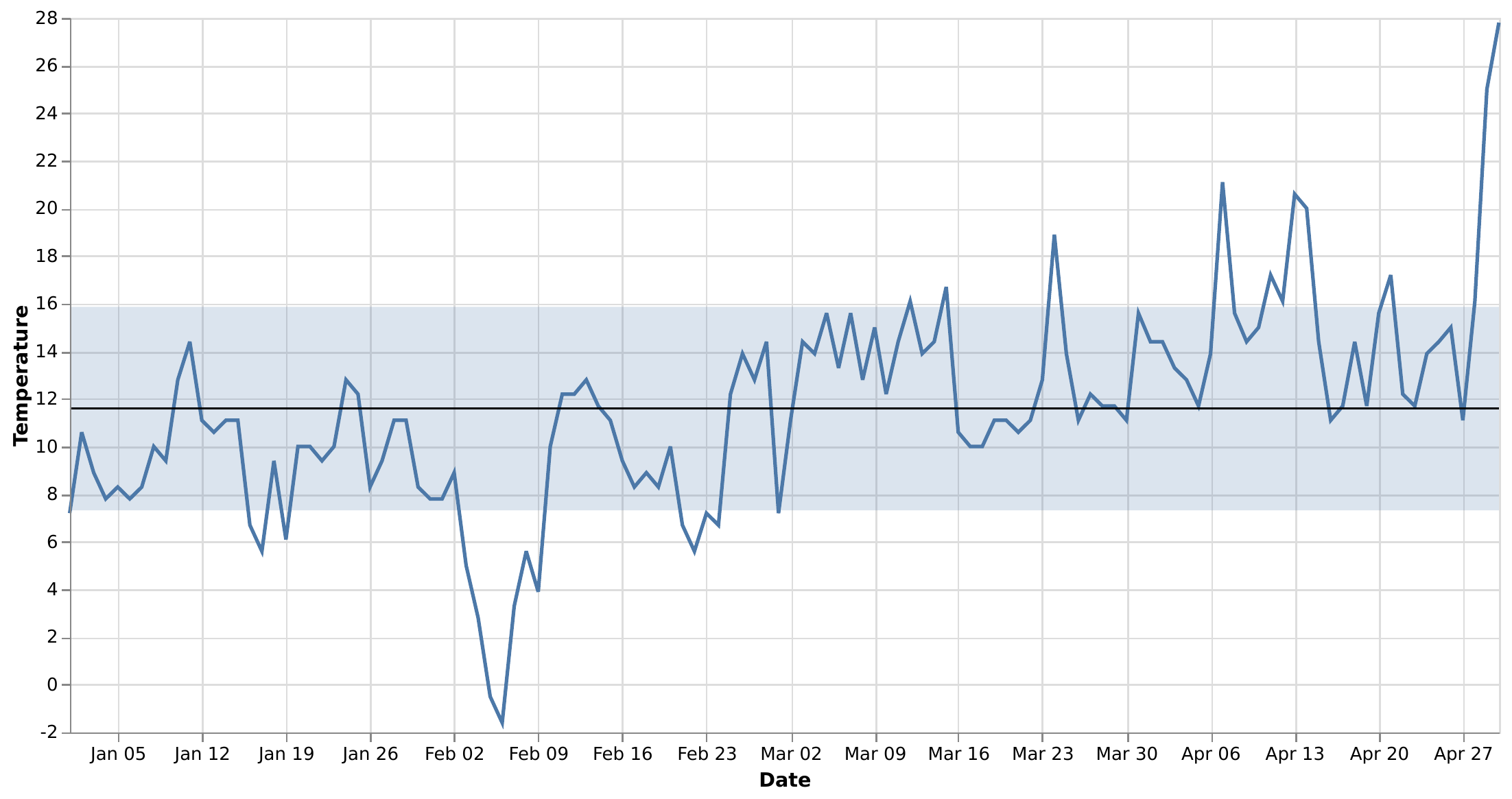}
\label{fig:spc}
}

\subfloat[Index charts compare to an indexed value rather than zero.]{
\includegraphics[width=0.7\columnwidth]{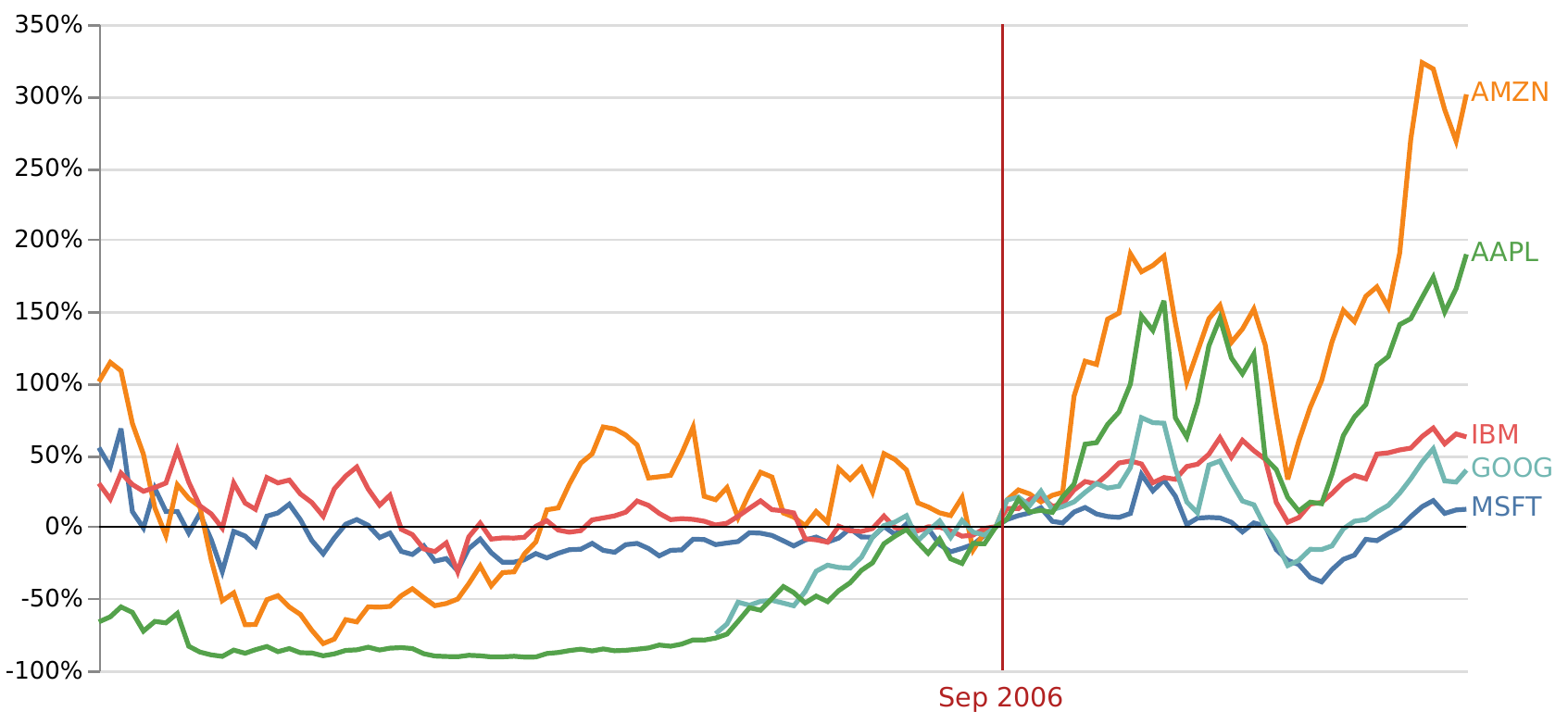}
\label{fig:index}
}

\subfloat[Stock charts must show small differences in stock value, as these can translate to enormous monetary gains or losses.]{
\includegraphics[width=0.7\columnwidth]{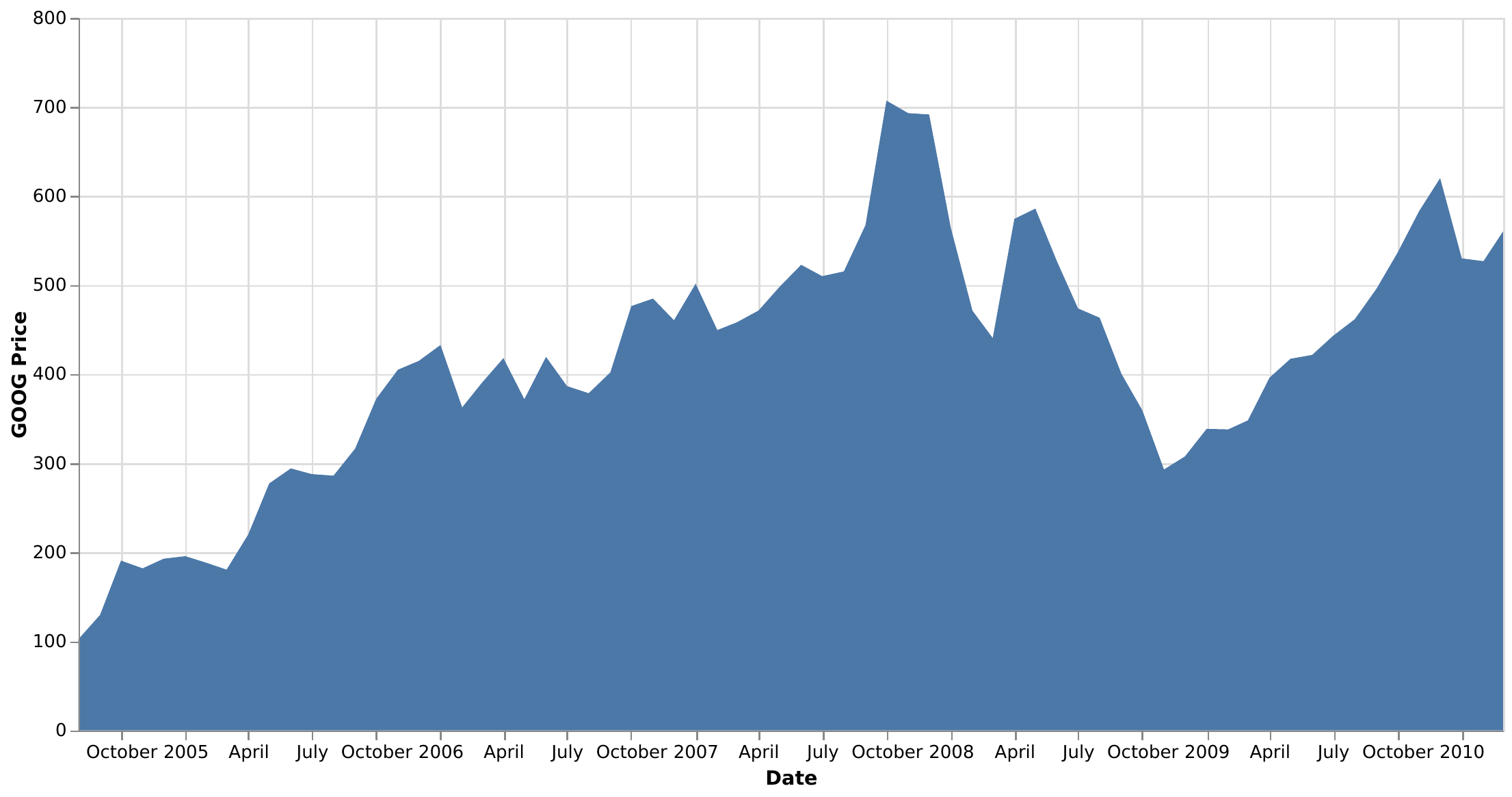}
\label{fig:stock}
}

\subfloat[Climate Anomaly charts rely on both highlighting deviation from a non-zero expected value but also emphasize the potentially disastrous impact of even minute changes in climate.]{
\includegraphics[width=0.7\columnwidth]{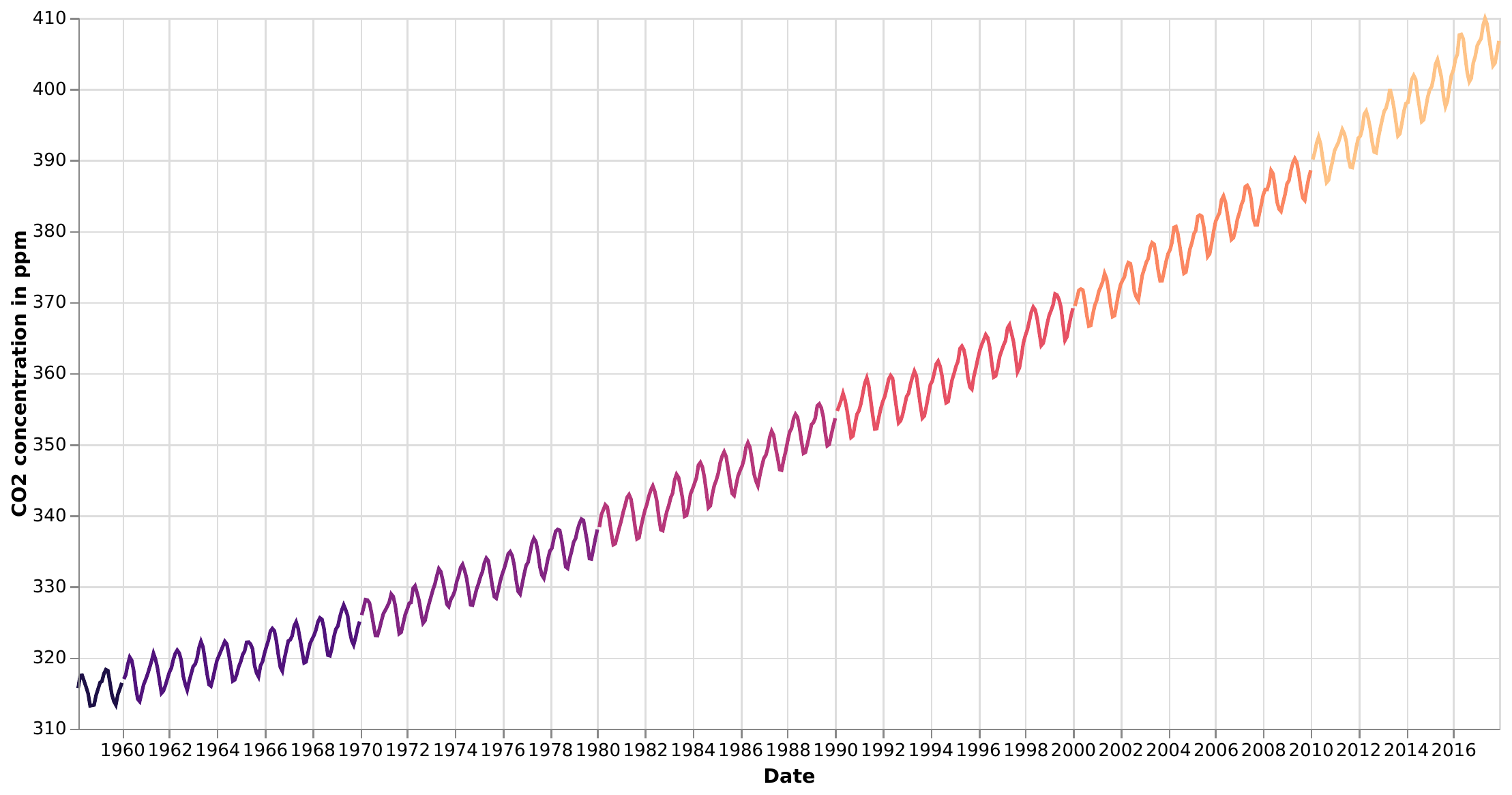}
\label{fig:anomaly}
}

\caption{The authors' reproductions of examples collected by Ben Jones\protect\cite{jonestweet}, of line charts where having a non-zero baseline for the y-axis is beneficial.}
\label{fig:goodtruncation}
\end{figure}

\begin{figure*}
    \centering
    \subfloat[Bar Chart]{
        \includegraphics[width=0.3\columnwidth]{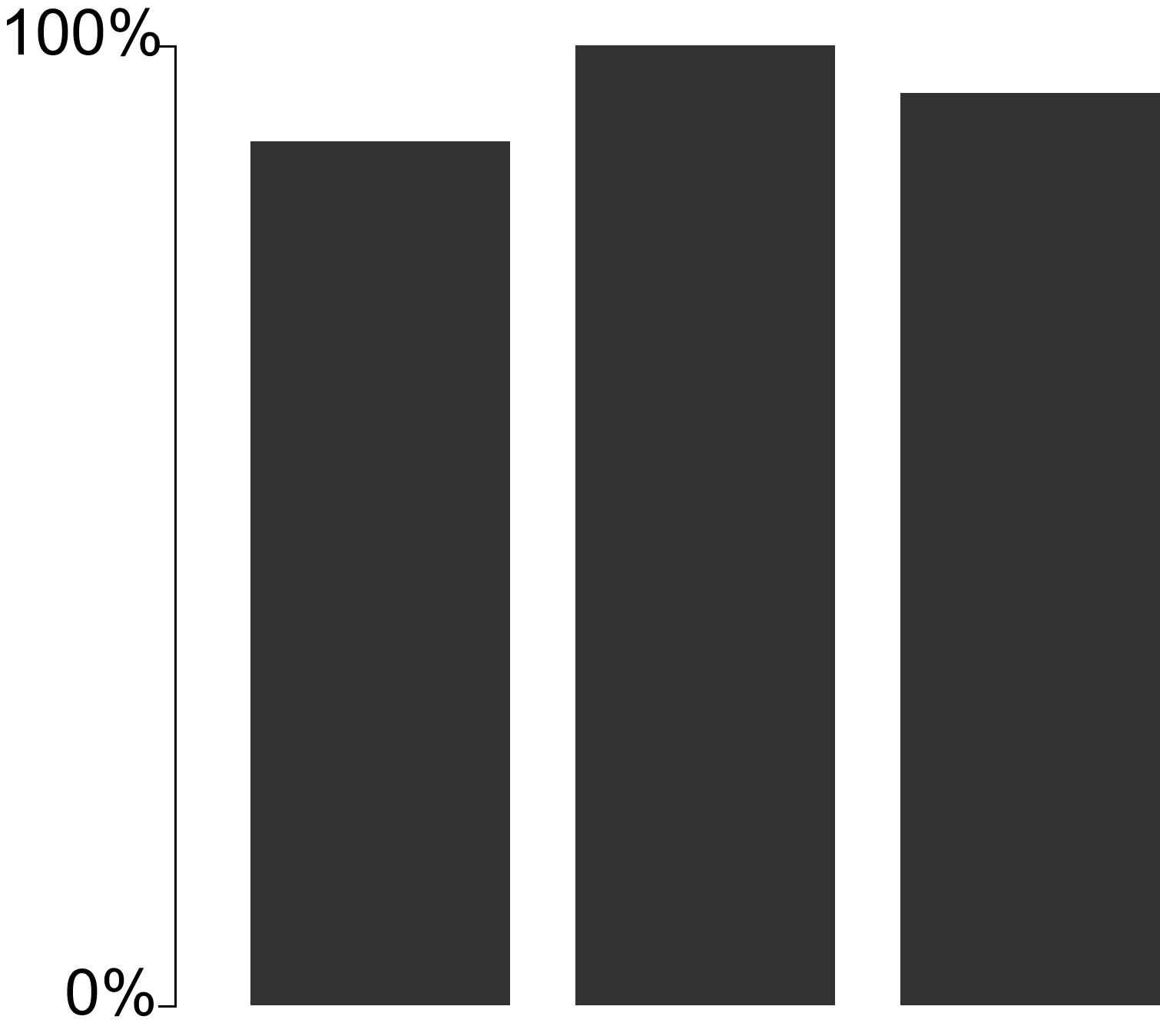}
        \label{fig:normalfix}
    }
    ~
    \subfloat[Broken Axes]{
        \includegraphics[width=0.3\columnwidth]{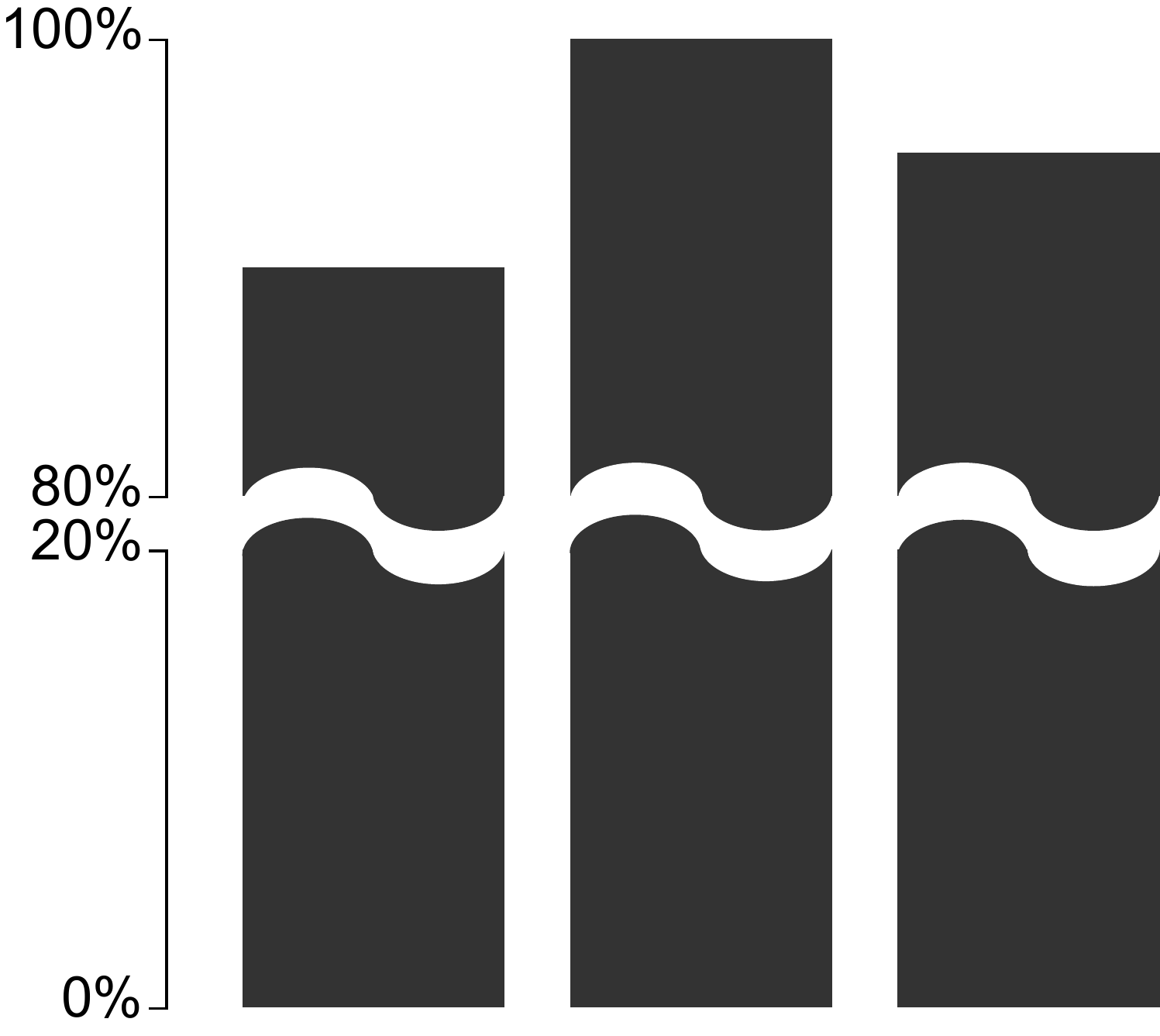}
        \label{fig:brokenfix}
    }
    ~
    \subfloat[Gradient Bar Chart]{
        \includegraphics[width=0.3\columnwidth]{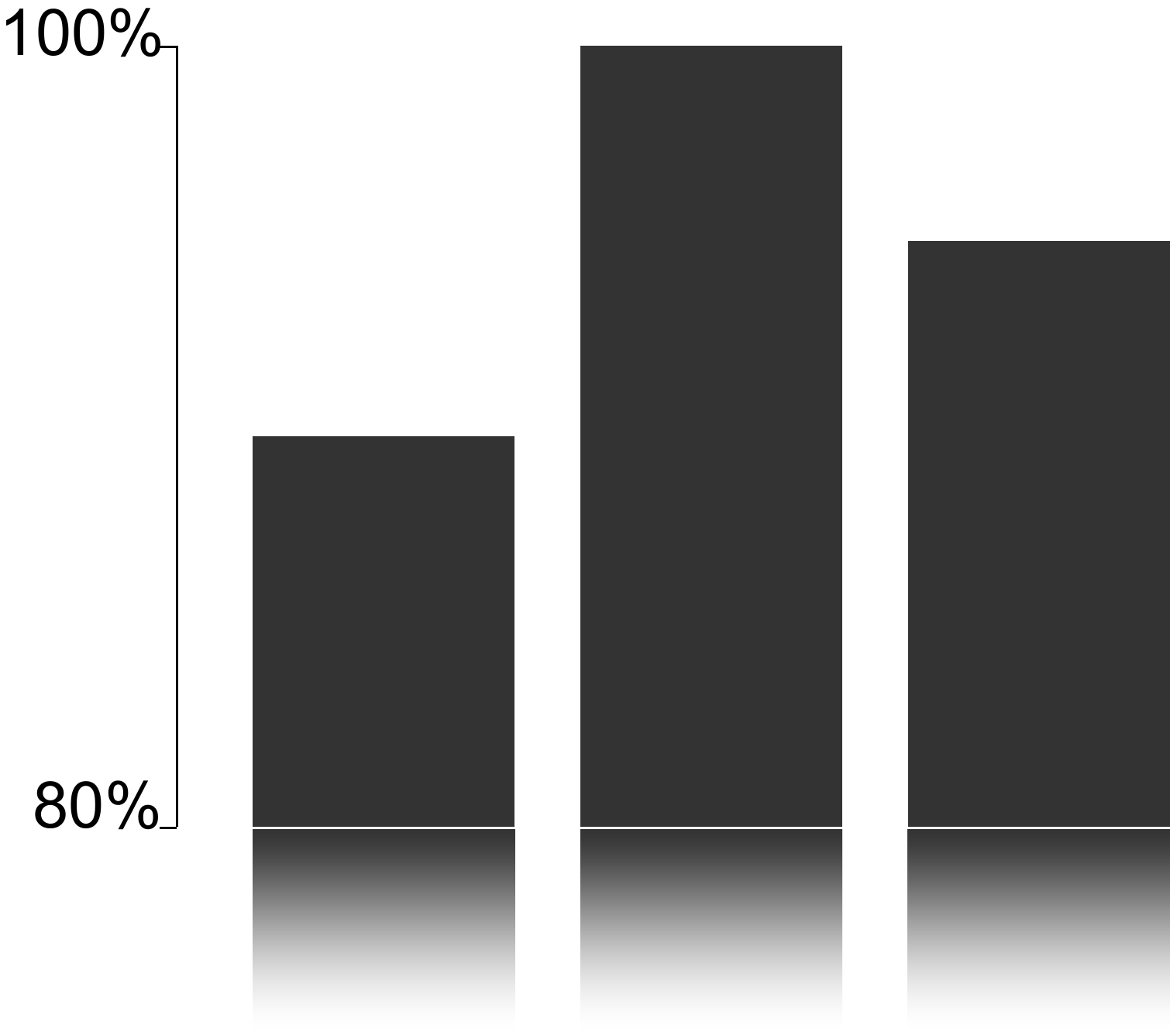}
        \label{fig:gradfix}
    }

    \subfloat[Torn Paper Chart]{
        \includegraphics[width=0.3\columnwidth]{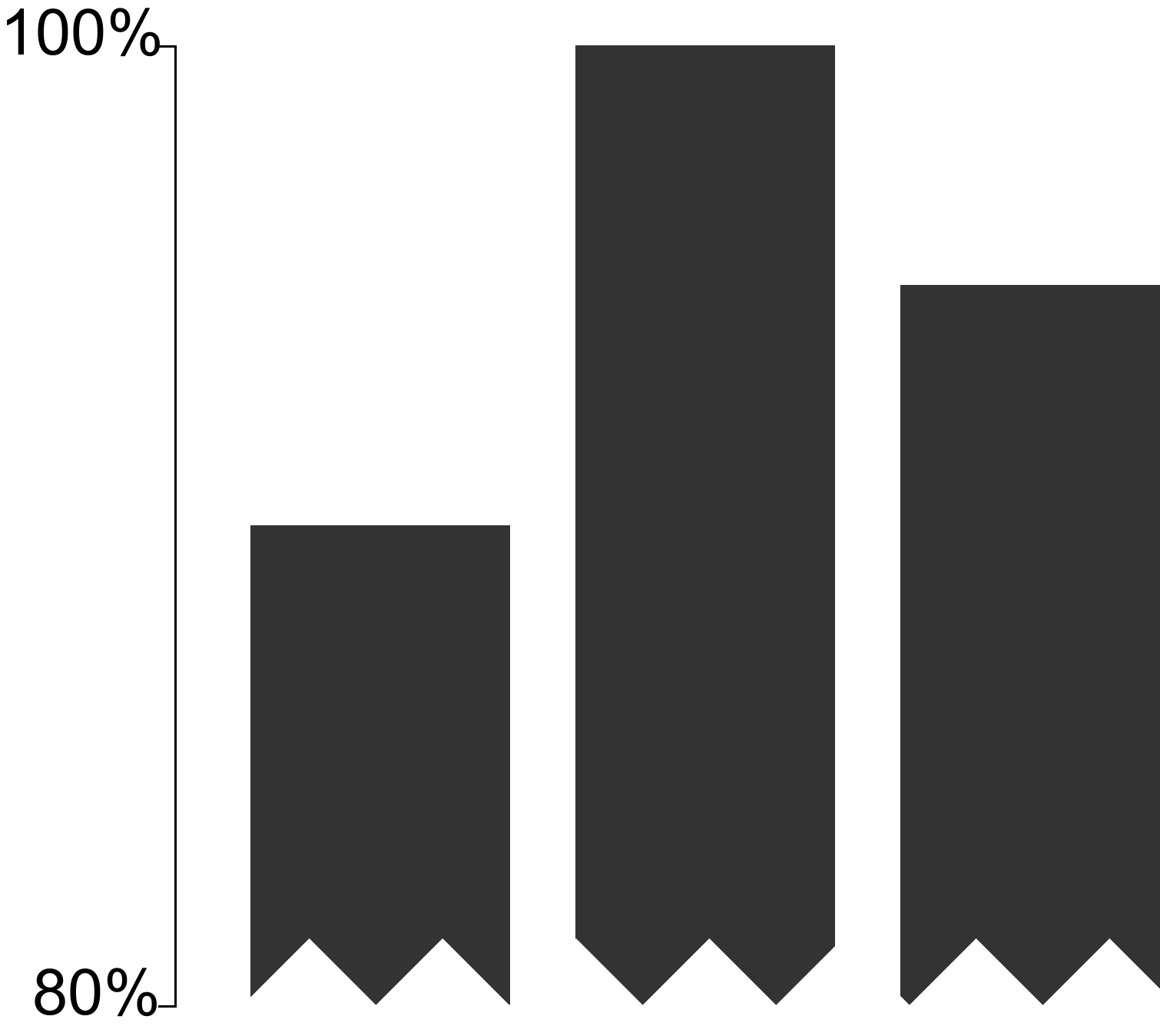}
        \label{fig:tornfix}
    }
    ~
     \subfloat[Panel Chart]{
        \includegraphics[width=0.3\columnwidth]{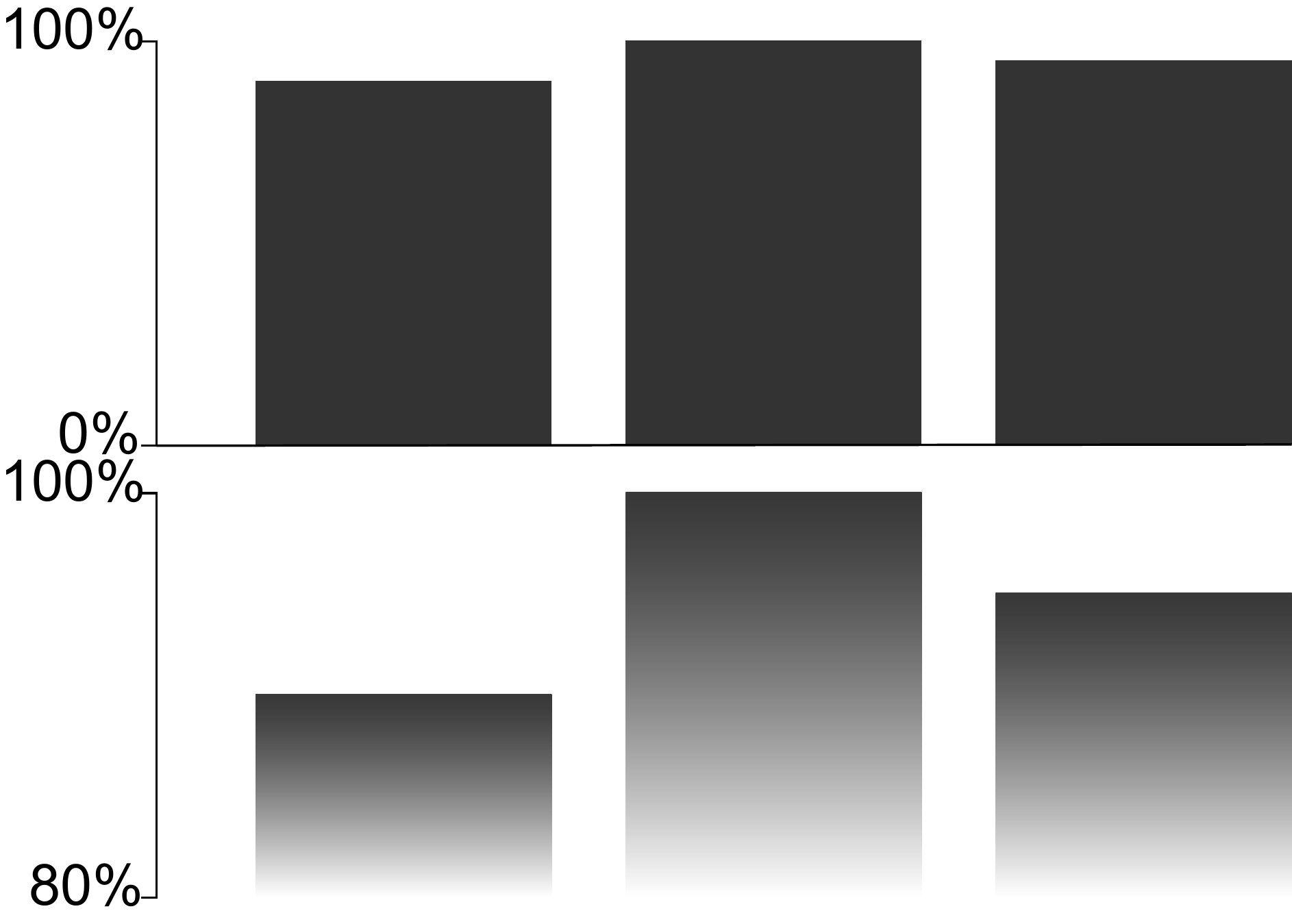}
        \label{fig:panelfix}
    }
    ~
    \subfloat[Interactive Focus+Context]{
        \includegraphics[width=0.5\columnwidth]{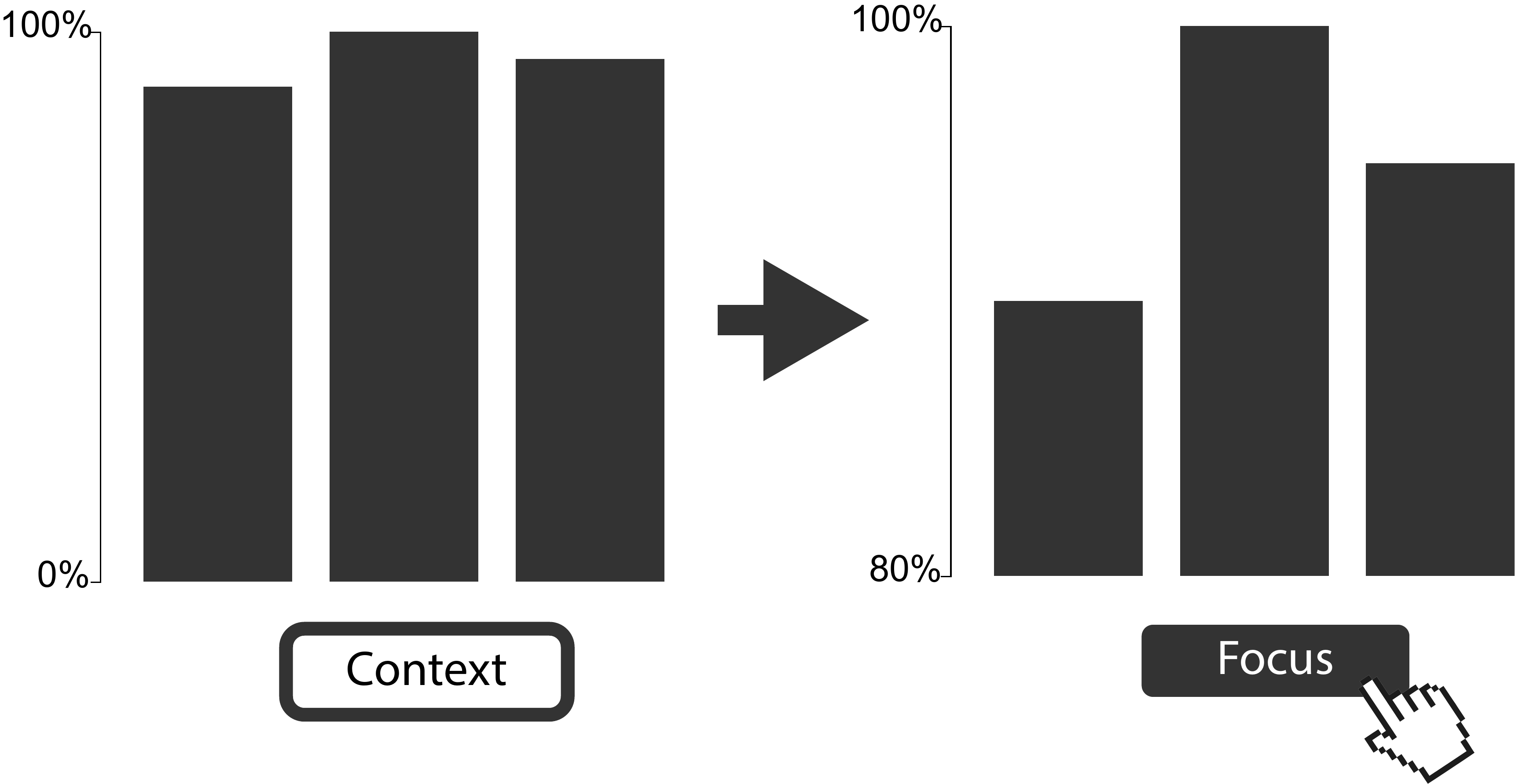}
        \label{fig:switchfix}
    }
    ~
    \subfloat[Bent Bar Chart]{
        \includegraphics[width=0.3\columnwidth]{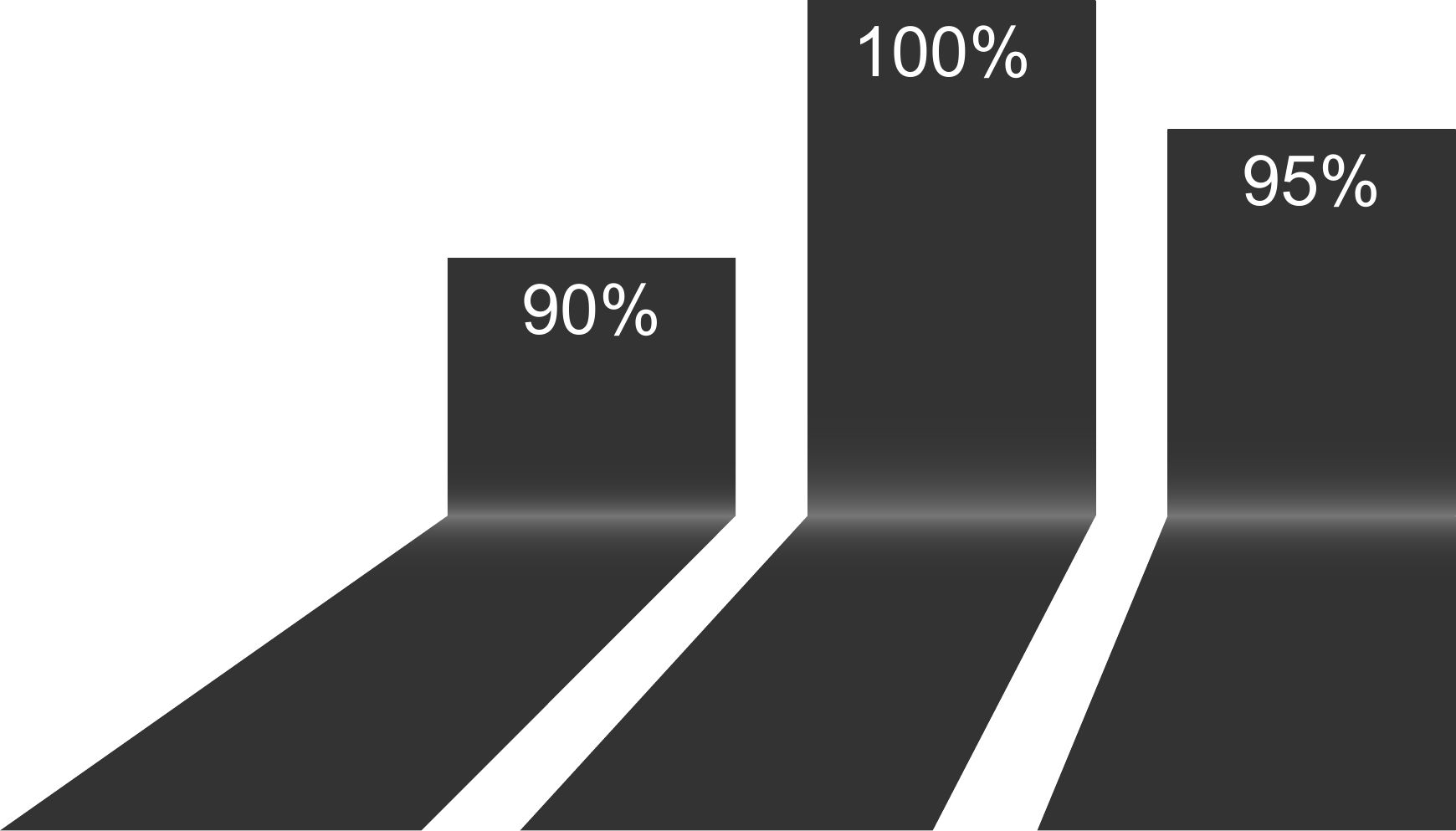}
        \label{fig:bendfix}
    }
    \caption{Proposed solutions for showing dynamic range in bar charts while indicating truncation or breakage of the y-axis. We test modifications of Figs.~\protect\ref{fig:brokenfix} and ~\protect\ref{fig:gradfix} in Experiments 2 and 3. Fig.~\protect\ref{fig:normalfix} is a bar chart with three large values in a narrow range. Brinton~\protect\cite{brinton1939graphic} recommends using a ``torn'' baseline to indicate an omitted $0$ as in Fig.~\protect\ref{fig:tornfix}. Pelier~\protect\cite{panelchart} claims that breaking the axis as in Fig.~\protect\ref{fig:brokenfix} is ``a bad idea'' and recommends ``Panel Charts'' as in Fig.~\protect\ref{fig:panelfix}, which (like Fig~\protect\ref{fig:gradfix}) uses a gradient to indicate values out of the current scale, but with a separate inset on top showing the full range of values. Ritchie et al.~\protect\cite{ritchie2019}, with similar reasoning, uses interaction to animate a transition from focus to context as in Fig.~\protect\ref{fig:switchfix}. Finally, Kosara~\protect\cite{eagereyes} suggests that ``bent'' 3D bar charts, as in Fig.~\protect\ref{fig:bendfix}, make accurate decoding difficult enough that the y-axis truncation is not particularly harmful-- the relative rankings of the categories is preserved even in 3D, and to accurately compare values viewers will likely need to consult the labels in any case. }
    \label{fig:fixes}
\end{figure*}

However, Chad Skelton believed that line graphs should not be a special category of exemption from truncation guidelines, and proposed the following~\cite{skeltonpost}:
\begin{quote}
Most line charts should start at zero. BUT not using baseline zero is OK if:\\
a) Zero on your scale is completely arbitrary (ie. temperature) OR\\
b) A small, but important, change is difficult or impossible to see using baseline zero.
\end{quote}

Related to these more permissive guidelines, Ben Jones specifically collected examples of common genres of line charts where non-zero baselines are not only accepted but are integral parts of the message of the chart (see Fig.~\ref{fig:goodtruncation}). These charts are examples where the analytical goals entail highlighting change from some baseline other than $0$, and in fact would mislead or confuse the viewer if they had non-truncated axes. These examples suggest that, for line charts at least, a $0$-baseline is not always appropriate.

Moreover, existing guidelines for the design of line graphs and scatterplots focus on making the overall trend as visible (and decodable with the least error) as possible~\cite{cleveland1988shape,heer2006multi,talbot2011arc,wang2018line}. These optimizations to chart aspect ratio typically assume that the chart covers the range of the data, rather than necessarily beginning at 0.

There are also competing design considerations to consider, across both encodings and  visual metaphors. The visual metaphors in charts impact how data are structured and interpreted~\cite{zacks1999bars,ziemkiewicz2008shaping,ziemkiewicz2009preconceptions}. Y-axis truncation breaks the visual conventions of bar charts, as the relative ratio of heights between two bars is no longer proportional to their difference in value (a bar that is twice as high may not represent a value that is twice as large). However, simply representing the same data as a line chart may not resolve this broken convention. Line charts have a \emph{continuous} encoding of position on the x-axis, and so employ a metaphor of \emph{continuity}. For data with discrete categories on the x-axis, a line chart may therefore be inappropriate. 
%Yet, in many cases data divided into discrete categories have (practical) significant effect sizes that occur in a narrow dynamic ranges (for instance, a difference of a few millimeters might mean a lot for data about a machine parts shop, but not for altimeter data from an aircraft). 
Highlighting changes in narrow ranges of a bar chart when the y-axis starts at 0 is challenging, and many existing solutions (such as those in Fig.~\ref{fig:fixes}) have not been empirically vetted. Designers may not have an ideal solution to the problem of depicting differences in small dynamic ranges of categorical data without violating some expectation or encoding practice in their visualization. 

Many existing guidelines in visualization lack rigorous empirical basis~\cite{kosara2016empire}. The results of graphical perception studies specifically on the impact of axes on graphical perception tasks are somewhat mixed. Pandey et. al~\cite{pandey2015deceptive} find that people rate differences in data as subjectively being greater in bar charts when the y-axis does not begin at 0.  Witt~\cite{witt2019graph}, by contrast tested truncation on both line charts and bar charts and find the least error in categorizing effect sizes when the range of the y-axis is 1.5 standard deviations. Witt also does not report any significant differences in truncation between visualization types.

While not examining truncation specifically, Berger~\cite{berger2005slippery} finds that line charts with higher slopes are perceived as more threatening than those with shallower slopes. As line graphs with truncated axes have steeper slope than those beginning at 0, this points to a potential bias in subjective assessment. However, Cleveland et al.~\cite{cleveland1982variables} find that \emph{expanding} scales (that is, making the axes extend far past the domain of the data, and as a consequence ``compressing'' the interior data distribution) results in higher estimates of correlation in scatterplots.

These empirical results point to two, interpretations of the impact of truncation. Pandey et al.~\cite{pandey2015deceptive} and Berger et al.~\cite{berger2005slippery} point to a ``bias'' in subjective perception of effects as a result of truncation. However, Witt~\cite{witt2019graph} and Cleveland et al.~\cite{cleveland1982variables} point to errors in estimation when the axes of a chart are extended past the range of the data.

\subsection{Research Questions}

In summary, truncating a y-axis may or may not always be dishonest, and so to be avoided. This anathema may or may not extend to line charts or time series data, or it could just be the case for bar charts. Even if line charts are an exception, whether it is permissible to truncate may depend on task (relative versus absolute change, trends versus values) and on data semantics (meaningful versus non-meaningful baselines). These ambiguities suggest the following questions for designers seeking to decide whether or not to truncate the y-axis in their chart:

\begin{itemize}
    \item Do the differences in visual design and framing behind line and bar charts result in different subjective effect sizes when the y-axis is altered? That is, \textbf{is the impact of y-axis truncation different between bar charts and line charts?}
    \item Does explicitly indicating that y-axis truncation has taken place (as in Fig.~\ref{fig:fixes}) reduce the bias introduced by truncation? That is, \textbf{can visual designs alleviate the exaggeration caused by truncation?}
\end{itemize}

These questions motivated our experiments on how people interpret the \emph{perceived severity} of effect sizes across visual designs, data metaphors, and indications of truncation.

\section{Experiment}
In order to assess how different visual presentations and analytical frames affect the inflation of perceived effect size introduced by truncating the y-axis, we conducted a series of two crowd-sourced experiments using the \url{prolific.ac} platform, approved by the IRB of Tableau Software. Experimental data from Prolific is comparable in quality to those from Amazon's Mechanical Turk platform~\cite{peer2017beyond}. The Prolific crowd-working platform is focused on studies rather than more general micro-tasks, and enforces minimum compensation rates of at least \$6.50/hour compared to the extremely low average earnings of workers on MTurk~\cite{hara2018data}. 

As per Huff, the major impact of y-axis truncation is not to misrepresent the values: ``nothing has been falsified;''~\cite{huff1993lie} rather, truncation inflates the \emph{subjective perception} of the rate of change of the values (Fig. \ref{fig:howtolie}). Therefore, while we investigate quantitative judgments in Experiment Three, the core of our experimental design is based on assessing \emph{qualitative} changes brought about by truncation. The design of our experiments was influenced by Pandey et al.~\cite{pandey2015deceptive}, who use a rating scale to assess the effect of various sorts of ``deceptive'' visualization practices, including y-axis truncation. Rating scales of statistical effect size have been in the context of assessing the impact of y-axis manipulations in Pandey et al. and Witt~\cite{witt2019graph}, but in general the use of rating scales to detect biases in statistical graphics have been used more widely, e.g., in Correll \& Gleicher~\cite{correll2014error}. Our central measure of interest was therefore the response to the 5-point rating item that related to how severe or important the differences in the data series were (the exact question text depended on the intended framing of the question). Higher ratings indicate a higher \emph{Perceived Severity} of the effect size. Our experimental design extends Pandey et al.'s work to a wider range of visual designs and task framings, and includes repeated within-subject trials on an array of different graphs (rather than just single exemplar pairs of deceptive and non-deceptive charts).

We also asked the participants if they thought the values were \emph{increasing} or \emph{decreasing} (in the trend framing), or if the first value was \emph{smaller} or \emph{larger} than the last value (in the values framing). We used this binary response as an engagement check and to test for comprehension of the chart data. Participants with unacceptably low accuracy at the engagement questions (more than three standard deviations lower than the mean performance) were excluded from analysis, but were compensated for their participation.

Since the main experimental measure was a subjective rating, we presented an initial set of 8 stimuli with every combination of \emph{slope} and \emph{truncation level} in order to present participants with the full range of visual effect sizes and so provide initial grounding for their ratings. These initial stimuli were discarded from analysis.

After the main rating task, we gave the participants a 13-item graphical literacy scale developed by Galesic and Garcia-Retamero~\cite{galesic2011graph}. Galesic and Garcia-Retamero reported a Cronbach's $\alpha$ of 0.85 for this scale, with some evidence of its utility as a cross-cultural measure of facility in interpreting charts and graphs. The scale does include one item that specifically tests for whether the participant noticed truncation of the y-axis. We collected the answer to this question separately, as well as collecting the overall scale value.

Lastly, we collected demographics data. In addition to standard items such as age and gender, we asked for three free-text responses, the first two of which were required to be non-empty:
\begin{itemize}
    \item What strategy or procedure did you use to complete the tasks? 
    \item Did you notice anything odd or unusual about the charts you saw during the task? 
    \item Any additional comments or feedback? 
\end{itemize}
Using these free text responses, a paper author and two third party researchers (one for experiments one and two, and another for experiment three) qualitatively coded whether or not the participants' free text responses specifically indicated that they noticed that the y-axes of some of the charts in the experiment were truncated. The coders then discussed mismatches ($\frac{5}{97}$ of codes), which were rectified into a final binary value.

Materials, data and analyses are available in our supplemental materials as well as \url{https://osf.io/gz98h/}.

\subsection{Experiment One: Framing Interventions}

\begin{figure}[t]
\centering
\subfloat[Bar Chart]{
\includegraphics[width=0.4\columnwidth]{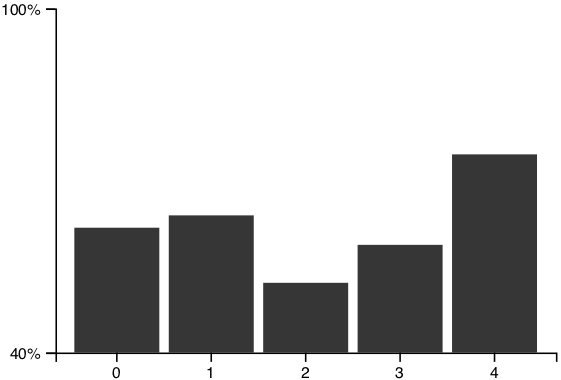}
\label{fig:exp1bar}
}
~
\subfloat[Line Chart]{
\includegraphics[width=0.4\columnwidth]{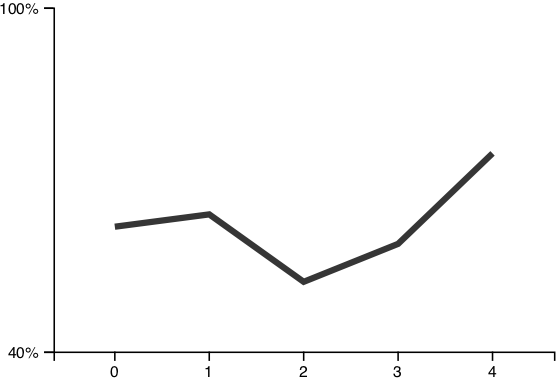}
\label{fig:exp1line}
}
\caption{The two visualization designs in Experiment One. In this example both have truncated axes.}
\label{fig:exp1cond}
\end{figure}

In unconstrained settings, line charts and bar charts can produce different sorts of judgments about the same backing data~\cite{zacks1999bars}.  Speculatively, one reason for this difference is that bar charts encourage the comparison of individual bars (and thus individual \emph{values}), whereas line charts encourage an assessment of the entire shape (and thus overall \emph{trend}). That is, the bar chart and line chart induce different \emph{framings} of the same data, based on their visual design and resulting \emph{visual metaphors}~\cite{ziemkiewicz2008shaping}. 

While truncation of the y-axes of equally sized line and bar charts both result in magnification of trend (and other measures such as correlation~\cite{cleveland1982variables}), we were interested in whether or not the differing framings associated with bar charts and line charts would result in differing impacts on judgments when the y-axis is truncated. If so, we were also interested if other strategies to promote different framings (such as text) could have this same impact, without the potential cost or inflexibility of switching visual designs.

\subsubsection{Methods}
As we were interested in a subjective measure with presumably wide person-to-person variance, we used a within-subjects design. To provide a wide variety of different visual effect sizes (in terms of differing bar heights or line slopes) while maintaining a relatively small number of trials to avoid fatigue, we had the following factors:
\begin{itemize}
    \item  \textbf{Visualization type} (2 levels): whether the data were visualized in a \emph{bar chart} or \emph{line graph}. See Fig. \ref{fig:exp1cond} for examples of these designs.
    \item \textbf{Question framing} (2 levels): either a \emph{value-based} or \emph{trend-based} task frame. For the \emph{value-based} framing, the engagement question was ``Which value is larger, the first value or the last value?'' and the effect size severity question was ``Subjectively, how different is the first value compared to the last value?'' with the labels ``Almost the Same,'' ``Somewhat Different,'' and ``Extremely Different'' for the first, third, and fifth items on our five-point rating scale. For the \emph{trend-based} framing, the engagement question was ``Are the values increasing or decreasing?'', the effect size severity question was ``Subjectively, how quickly are the values changing?'', and the rating labels were ``Barely,'' ``Somewhat,'' and ``Extremely Quickly.''
    \item \textbf{Truncation Level} (3 levels): where the y-axis of the visualization began: either at $0$, $25$, or $50\%$.
    \item \textbf{Slope} (2 levels) : How much increase (or decrease) there was between the first and last values in the data, either $12.5\%$ or $25\%$.
    \item \textbf{Data Size} (2 levels): whether there were \emph{two} or \emph{three} data values in the visualization. If there were three data values, the center value was at the midpoint of the first and last items, with a uniform random jitter between $[0,\frac{slope}{4}]$.
\end{itemize}

Participants saw one of each combination of factors, for a total of $2 \times 2 \times 3 \times 2 \times 2 = 48$ stimuli, presented in a randomized order. Whether the values were \emph{increasing} or \emph{decreasing} was an additional random factor.

\subsubsection{Hypotheses}

As with Pandey et al.~\cite{pandey2015deceptive} for bar charts, and Berger~\cite{berger2005slippery} for line charts, we expected that \textbf{charts with more axis truncation would be perceived as having more severe effect sizes than those with less truncation}.
 
While there has been some initial work on the impact of framing in visualization~\cite{hullman2011visualization,kong2018frames,xiong2017curse}, and how different visual metaphors can impact how we make use of information in charts~\cite{ziemkiewicz2008shaping,ziemkiewicz2009preconceptions}, there is relatively little empirical work on how different visual designs impact framing. Therefore our hypotheses about framing were weakly held. In particular, based on the misleading visual metaphor introduced by truncation in bar charts (where the relative size of the bars is not a proxy for the relative difference in values) we expected that \textbf{Bar charts would be more greatly impacted by truncation than line charts} in terms of amplifying perceived severity of effect sizes. Similarly, we believed that \textbf{the values framing would be more greatly impacted by truncation than the trend framing}, as comparison of individual values in a truncated graph is fraught and potentially misleading (especially if the axis legend is ignored).

\subsubsection{Results}
\begin{figure*}
    \centering
    \subfloat{
        \includegraphics[width=0.65\textwidth,valign=t]{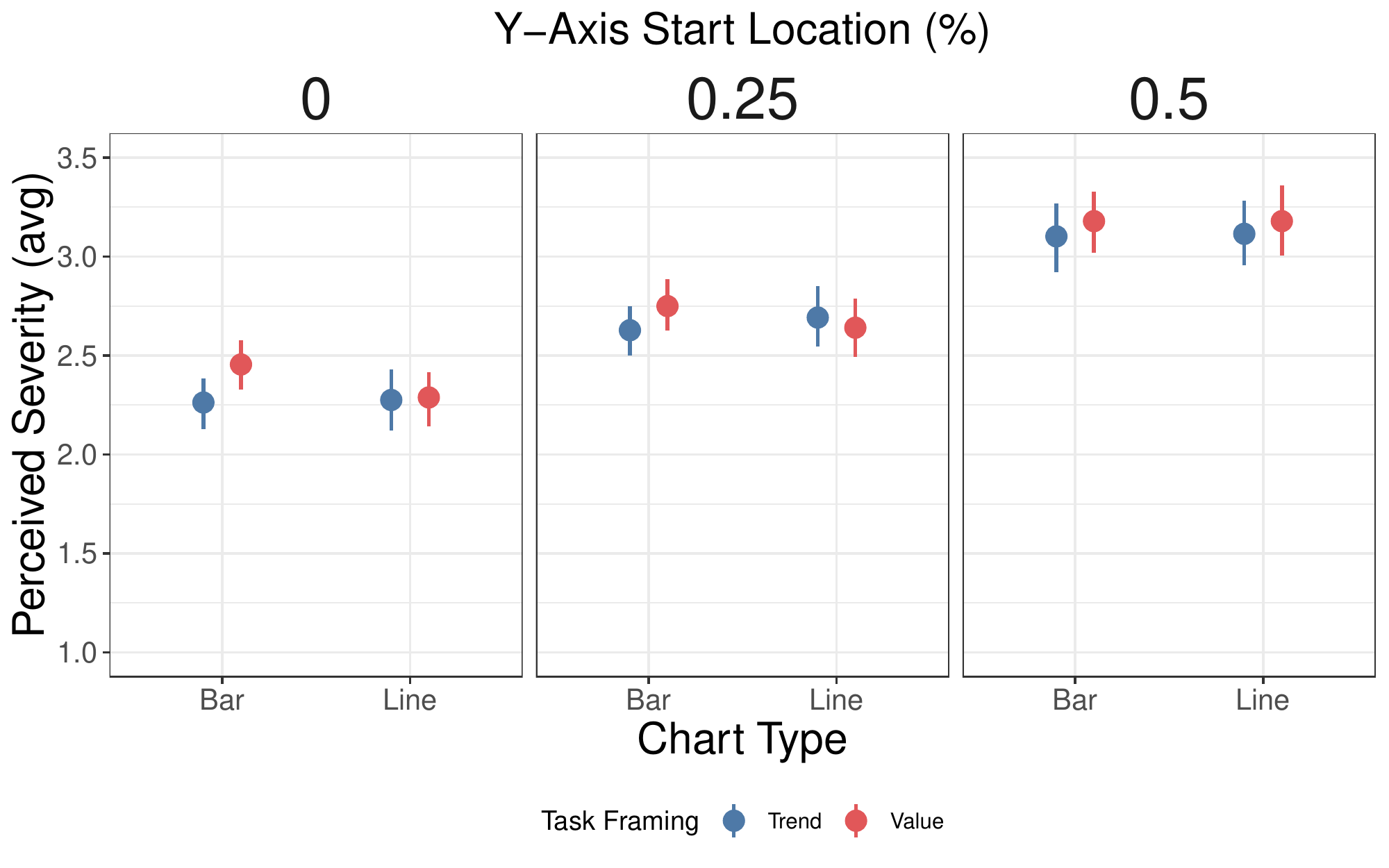}
        \label{fig:exp1All}
    }
    ~
    \subfloat{
        \includegraphics[width=0.3\textwidth,valign=t]{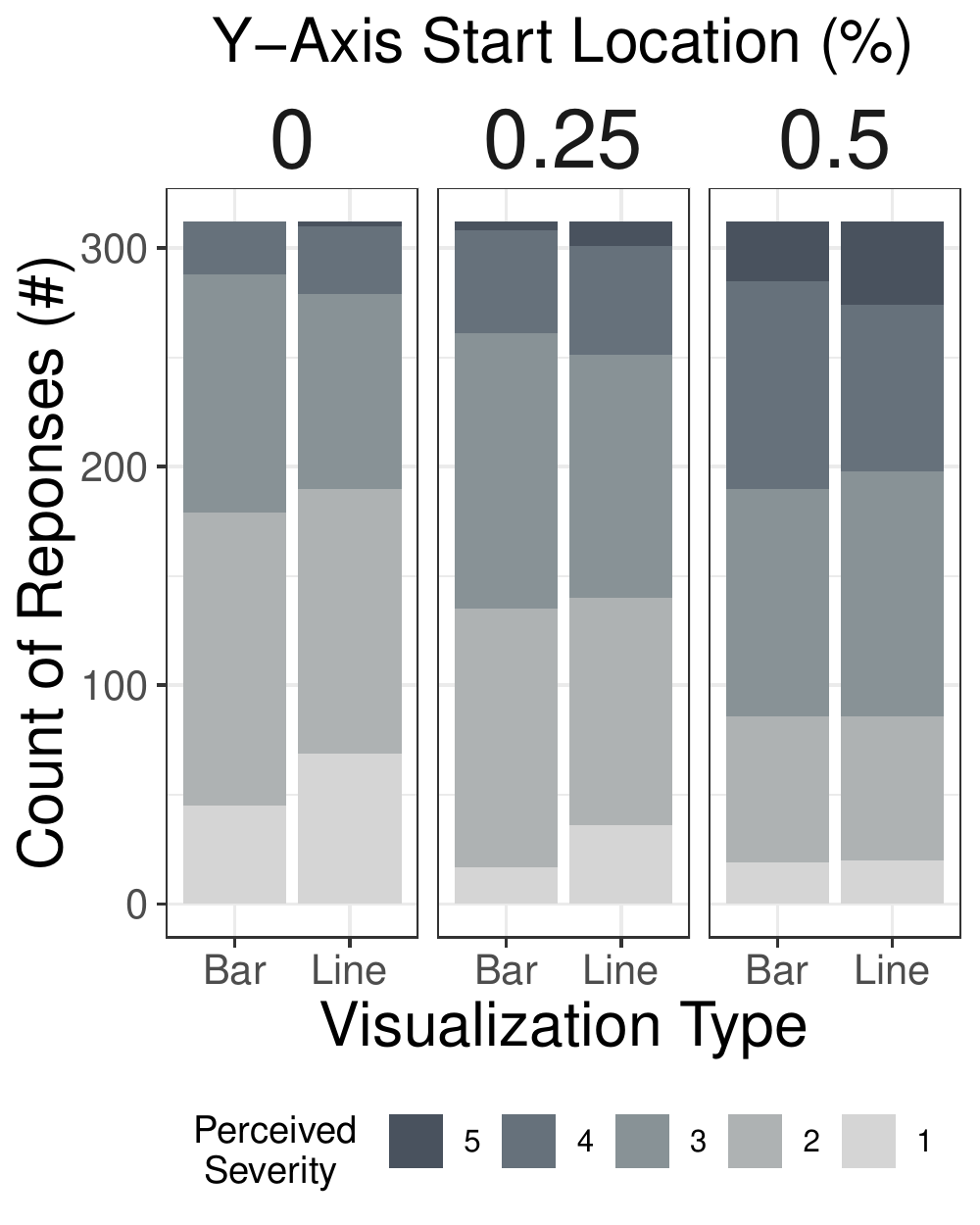}
        \label{fig:exp1Stacked}
    }
    
    \caption{Results from Experiment One. Increasing the starting point of the y-axis results in larger perceived severity in effect size. Neither the visual design (bar or line chart) nor the method of soliciting the perceived severity (focusing on either individual \emph{values} or overall \emph{trend}) produced significant differences in perception of effect size. Error bars represent 95\% bootstrapped confidence intervals of the mean. The figure on the right shows the raw counts of rating responses across visualization types.}
    \label{fig:exp1results}
\end{figure*}

We recruited 40 participants for this task (21 male, 18 female, 1 with a non-binary gender identity, $M_{age} = 27.7$, $SD_{age}=8.4$). We paid participants \$4 for this task, for an empirical effective hourly rate of \$12/hour. On average, participants scored well on the 13-item Galesic and Garcia-Retamero graphical literacy scale ($M_{correct}=10$, $SD_{correct}=2$). In particular, 31 (78\%) participants correctly answered the scale item associated with y-axis truncation. Additionally, $45\%$ of participants explicitly mentioned y-axis manipulation in their post-task free responses. Participants also scored highly on our engagement question, correctly labeling the direction of the effect ($M_{correct}=0.98$, $SD_{correct}=0.07$), with the exception of one participant, whose performance was more than three standard deviations from the mean ($62.5\%$). The data from this participant was excluded from our analysis.

We conducted a repeated measures ANOVA testing the effect of truncation level, visualization type, and question framing, and their interactions on perceived severity.

Our results support our first hypothesis: \textbf{increased y-axis truncation results in increased perceived severity} ($F(2,76)=89$, $p<0.0001$). A post-hoc pairwise t-test with a Bonferroni correction confirmed that the perceived severity of all three levels of truncation were significantly different from each other. Fig.~\ref{fig:exp1results} illustrates this result, broken out by visualization type.

Our results fail to support our second hypothesis. \textbf{There was no significant effect of visualization design on perceived effect size} ($F(1,38)=0.5$, $p=0.50$). Fig.~\ref{fig:exp1results} shows similar responses to different visualizations across all levels of truncation.

Our results only weakly support our last hypothesis. While there \emph{was} a significant effect of framing on perceived effect size (($F(1,38)=7.4$, $p=0.01$), a post-hoc pairwise t-test with a Bonferroni correction did not find a significant difference between the value and trend question framings. Additionally, this effect was quite small: an average decrease in perceived severity of $0.07$ for responses using the \emph{trend} framing (based on a $1-5$ rating scale), compared to an increase of $0.36$ for starting the y-axis at $25\%$ rather than $0\%$. Fig.~\ref{fig:exp1results} shows this result, broken out by visualization design.

%\begin{figure}
%    \centering
%    \includegraphics[width=0.8\columnwidth]{exp1Noticed.pdf}
%    \caption{Results from Experiment One. People who explicitly mentioned the truncated or potentially misleading y-axes reported less severity in effect sizes. Results were similar for Experiment Two. Error bars represent 95\% bootstrapped confidence intervals of the mean.}
%    \label{fig:exp1noticed}
%\end{figure}

\subsection{Experiment Two: Visual Design Interventions}

The results of our first experiment show no robust difference in the impact of truncation on bar charts and line charts: truncation results in largely qualitatively assessed effect sizes in both types of graphs. This result suggests that designers may have to employ other methods to indicate that a y-axis has been truncated. A common solution to this problem is to employ the visual metaphor of the ``broken'' or ``continued'' axis. Wikipedia recommends indicating truncated axes with glyphs~\cite{wikimislead} that convey a break from 0 to the start of the truncated axis. To our knowledge, there is no empirical work on whether or not these indications of breaks alter judgments about values. As such, we performed an experiment with similar methodology to our first experiment in order to assess the impact of visual design elements in bar charts that indicate truncation or continuation on perceived effect size.

\begin{figure*}[t]
\centering
\subfloat[Bar Chart]{
\includegraphics[width=0.4\columnwidth]{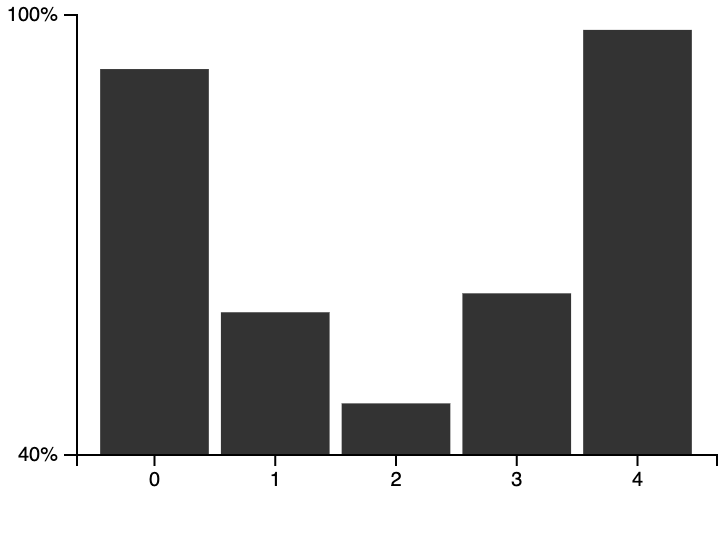}
\label{fig:exp2bar}
}
~
\subfloat[Bar Chart with Broken Axis]{
\includegraphics[width=0.4\columnwidth]{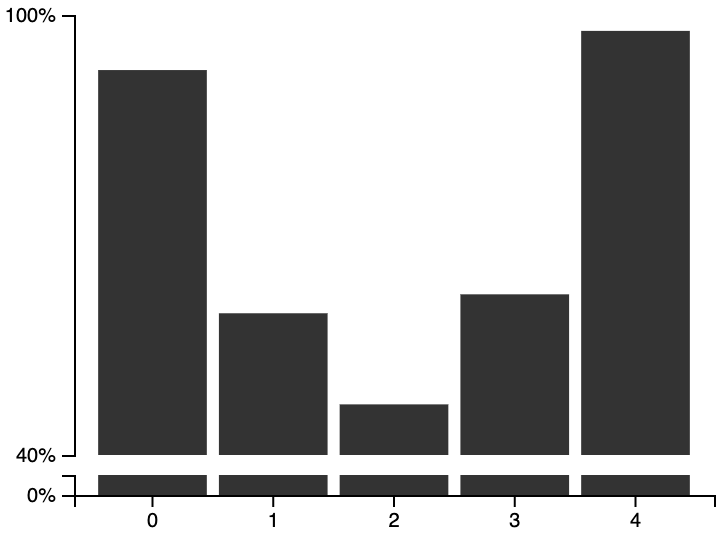}
\label{fig:exp2brokenbar}
}
~
\subfloat[Bar Chart with Gradient]{
\includegraphics[width=0.4\columnwidth]{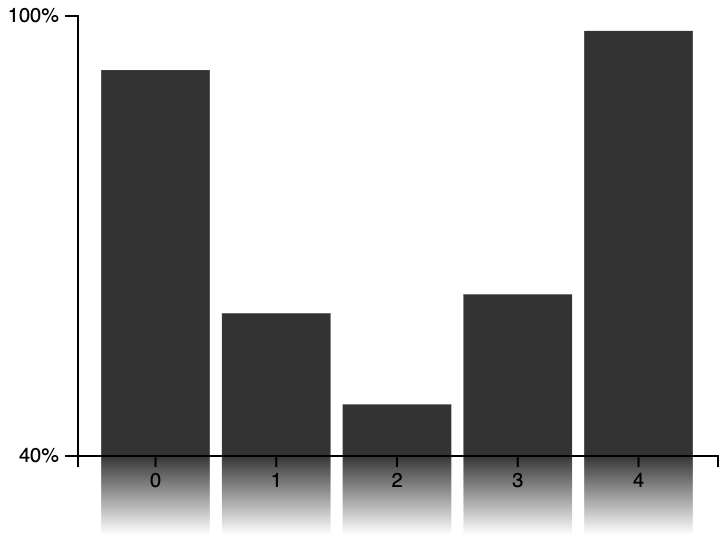}
\label{fig:exp2gradbar}
}

\caption{The three visualization designs in Experiment Two and Three. A broken bar and axis legend indicates truncation in Fig.~\ref{fig:exp2brokenbar}, whereas the continuation of the axis beyond the chart is indicated by a gradient in Fig.~\ref{fig:exp2gradbar}. When there is no truncation, breaking the axis and indicating that the bars continue is inappropriate. In those cases, the alternative designs devolve into traditional bar charts, as in Fig.~\ref{fig:exp2bar}.}
\label{fig:exp2cond}
\end{figure*}

\subsubsection{Methods}
Our results from Experiment One were initial evidence that subtle framing effects were not sufficient to reliably impact estimations. As such, we excluded that factor, sticking with the trend framing from the first experiment. Instead, we focused on a narrower set of designs that have been proposed to ameliorate the impact of y-axis truncation in bar charts.

There are many possible designs that have been proposed to indicate y-axis breaks (see \ref{fig:fixes}). However, we were interested in cases where there is an indication of an axis break \emph{per se}, without any additional alteration of the (truncated) height of the bars, screen space dedicated to the chart, or additional required user interaction, limiting us to a subset of the solutions proposed in prior work.

We therefore modified two of the proposed fixes as exemplars of designs where the truncation of the y-axis is not only visible in reading the y-axis labels, but is an integral component of the visual metaphor of the chart (see Fig. \ref{fig:exp2cond}). \emph{Bar charts with broken axes} are a common choice to indicate a truncated axis. In Fig. \ref{fig:exp2brokenbar} we use a broken axis design with both a break on the y-axis as well as a break in the bars themselves to reinforce the metaphor across the entire chart. Bar charts with irregular shapes on the bottom have been reported as complicating decoding~\cite{skau2015evaluation}, so we use rectangular glyphs to indicate breakage rather than the ``wavy'' or ``jagged'' glyphs commonly uses to indicate breaks~\cite{wikimislead}(as in Fig. ~\ref{fig:brokenfix}). Our second design, a \emph{bar chart with a gradient bottom} (Fig, \ref{fig:exp2gradbar}), has been empirically considered for other scenarios by both Skau et al. ~\cite{skau2015evaluation} and Diaz et al.~\cite{diaz2018improving}. Skau et al. in particular were investigating the case where the gradient is an artistic embellishment (mean to convey e.g., sitting on a reflective surface) rather than conveying continuation. They found that the gradient caused overestimation of value in their task setting, which is potentially advantageous for our task (where visual exaggeration of value might counteract the effect of truncation). For both conditions, the heights of the bars above the break (for the broken axis chart) and above the y-axis (for the gradient bottom chart) were equal to the height of the bars in the standard truncated bar chart. The pre-break axes and under-bar gradients were therefore constrained to a narrow area underneath the chart. We include code for additional potential designs in our study materials.

For this experiment, reused the factors and factor levels from the previous experiment, but used only the \emph{trend-based} question framing, and new visualizations:
\begin{itemize}
    \item  \textbf{Visualization type} (3 levels): whether the data was visualized in a \emph{bar chart}, \emph{broken axis bar chart}, or \emph{gradient bar chart}. See Fig. \ref{fig:exp2cond} for examples of these designs.
\end{itemize}

Participants saw one of each combination of factors, for a total of $3 \times 3 \times 2 \times 2 = 36$ stimuli, in a random order. As with the previous experiment, we included an initial set of $8$ stimuli illustrating the full range of effect sizes in order to assist in calibrating the participants' subjective judgments, for a total of $44$ total stimuli, but these calibration stimuli were excluded from analysis.

\subsubsection{Hypotheses}

We had only a single hypothesis for this experiment: \textbf{visual designs with non-zero axes that indicate y-axis breaks or continuations would be perceived as having smaller effect sizes than standard bar charts}. We believed that these visual indications would make the truncation harder to ignore or overlook, and promote caution or reflection in judgments.

\subsubsection{Results}
\begin{figure*}
    \centering
    \subfloat{
        \includegraphics[width=0.6\textwidth]{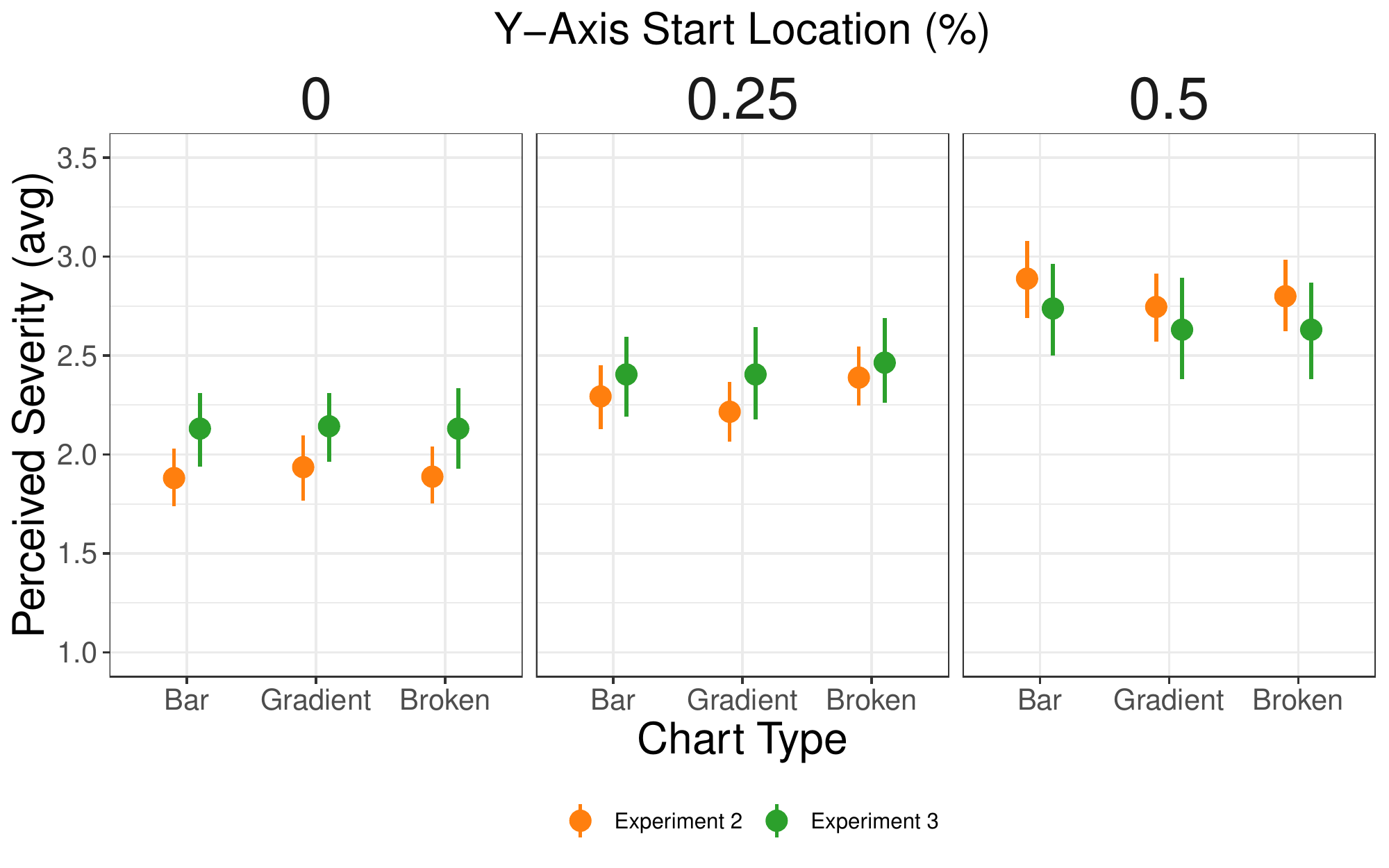}
    }
    ~
    \subfloat{
    \includegraphics[width=0.35\textwidth]{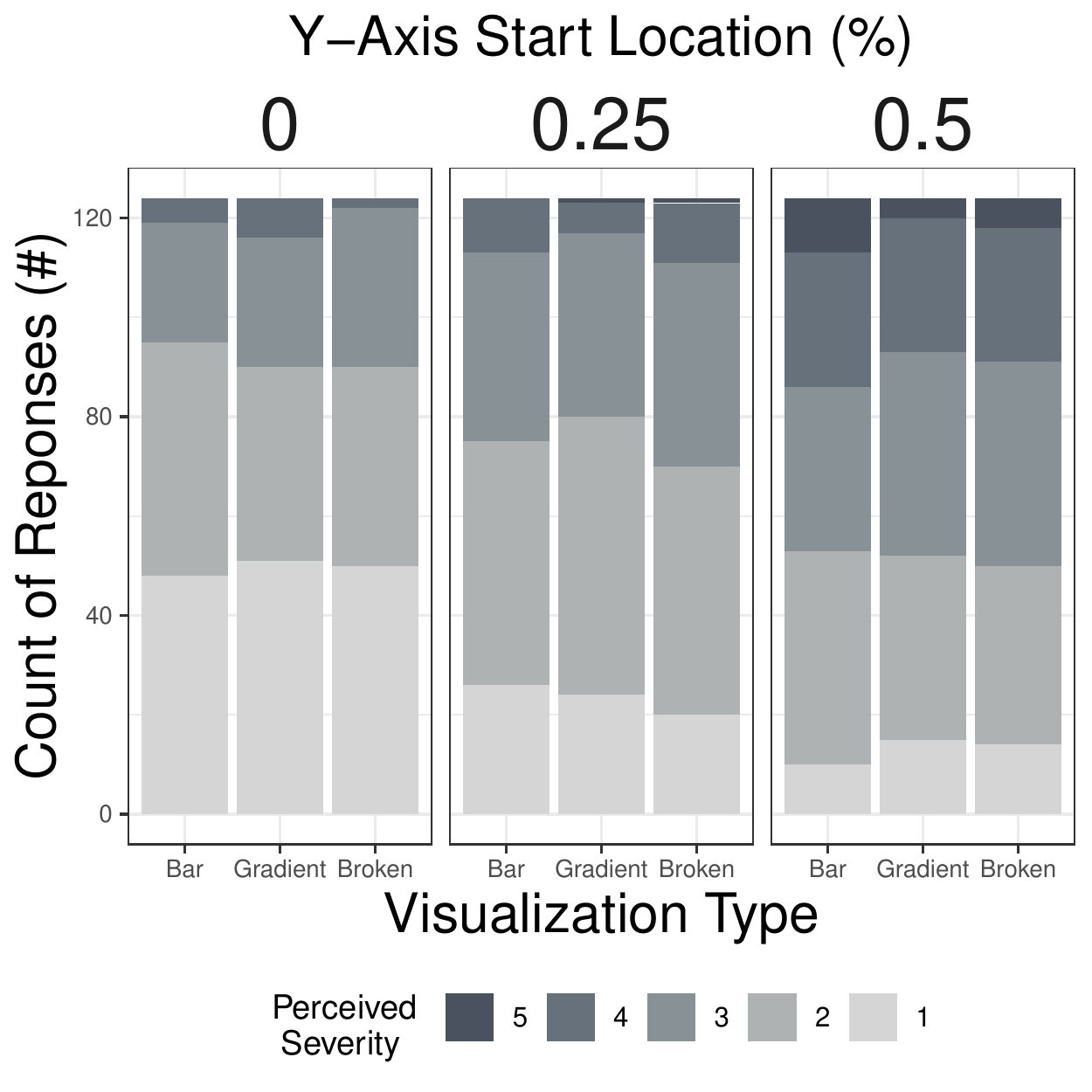}
    }
    \caption{Results from Experiments Two and Three. While broken axes may \emph{indicate} that a y-axis is truncated, and a gradient fill may \emph{connote} that the bars extend beyond the visualized chart area, neither intervention had a consistent impact on perceived severity; increased axis truncation resulted in similar increases in perceived severity. Error bars represent 95\% bootstrapped confidence intervals of the mean. Note that when there is no truncation (the y-axis begins at $0\%$), all three designs were visually identical. Error bars represent 95\% bootstrapped confidence intervals of the mean. The figure on the right shows the raw counts of rating responses across visualization types for Experiment Two-- results across both Experiments Two and Three were similar.}
    \label{fig:exp2designs}
\end{figure*}

We recruited 32 participants for this task (20 female, 12 male, $M_{age} = 29.0$, $SD_{age}=11.7$). We paid participants \$4 for this task, for an empirical effective hourly rate of \$16/hour. Participants scored well on the Galesic and Garcia-Retamero graphical literacy scale ($M_{correct}=10.8$, $SD_{correct}=2$). Similar to the previous study, $25$ ($78\%$) participants correctly answered the scale item connected with y-axis truncation. Participants also scored highly on our engagement question, correctly labeling the direction of the effect ($M_{correct}=0.99$, $SD_{correct}=0.02$). One participant had performance more than three standard deviations from the mean ($93.2\%$). The data from this participant was excluded from our analysis.

We conducted a repeated measures ANOVA testing the effect of truncation level, visualization type, data size, and their interactions on perceived severity. We built our model on the subset of trials where the truncation level was $>0$, as those were the trials with visual differences between designs. 

Our results fail to support our first hypothesis: \textbf{there was no significant difference between perceived severity among visualization designs} ($F(2,60)=3.1$, $p=0.05$). A post-hoc pairwise t-test with a Bonferroni correction failed to find any significant difference between visualization designs. Fig. \ref{fig:exp2designs} illustrates the performance of all three designs across different truncation levels.

We only coded $31\%$ of participants as having specifically mentioned y-axis truncation in their post-task free text responses, compared to $47\%$ in the first experiment. It is possible that the alternative conditions made the truncation of the axis so ``obvious'' that it was not felt necessary to comment upon, but we had no specific hypothesis to this effect.

\subsection{Experiment Three: Bias in Value Estimation}
    \begin{figure*}[ht!]
    \centering
    \subfloat[]{
    \includegraphics[width=0.85\columnwidth]{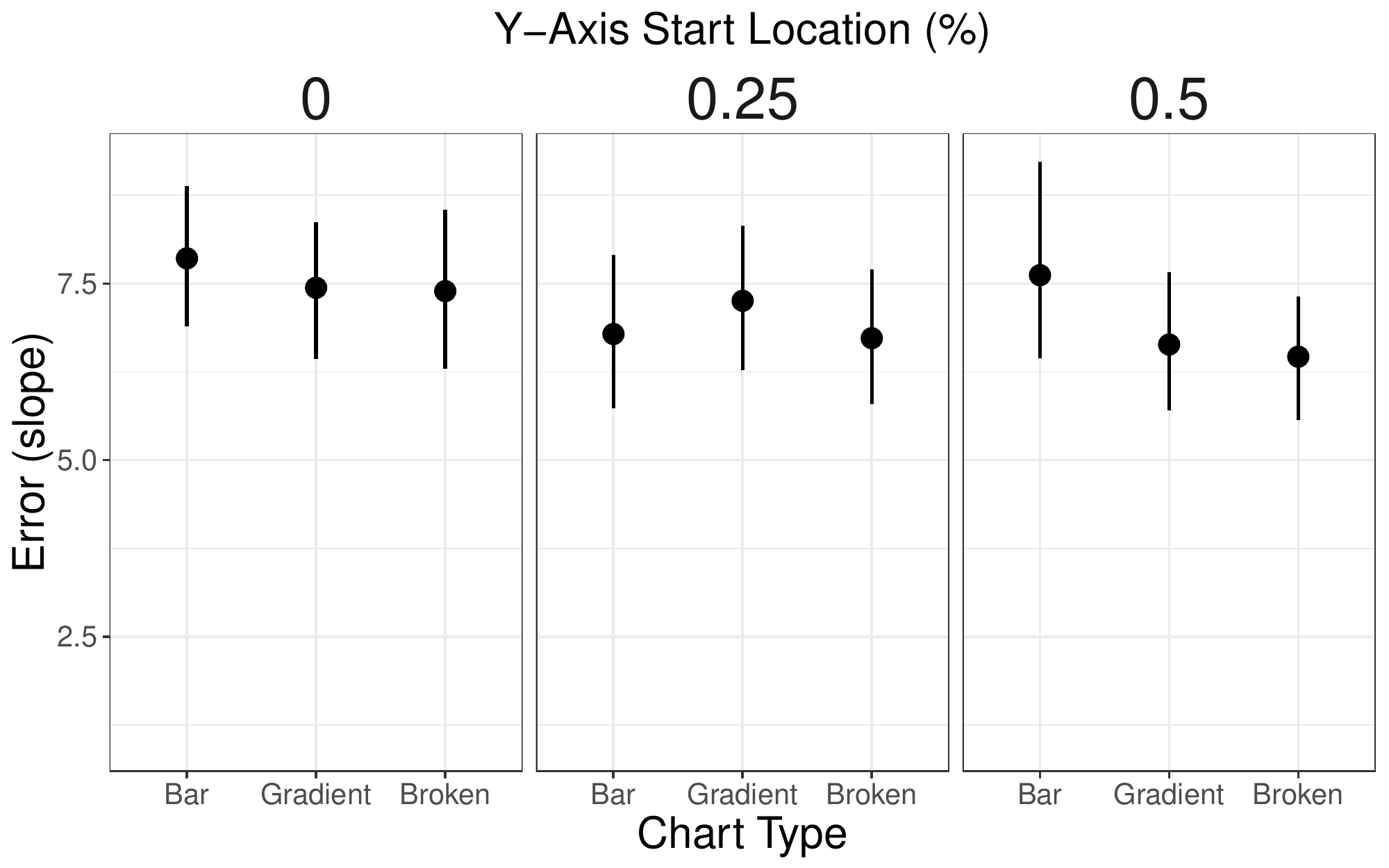}
    \label{fig:exp3trend}
    }
    ~
    \subfloat[]{
    \includegraphics[width=0.85\columnwidth]{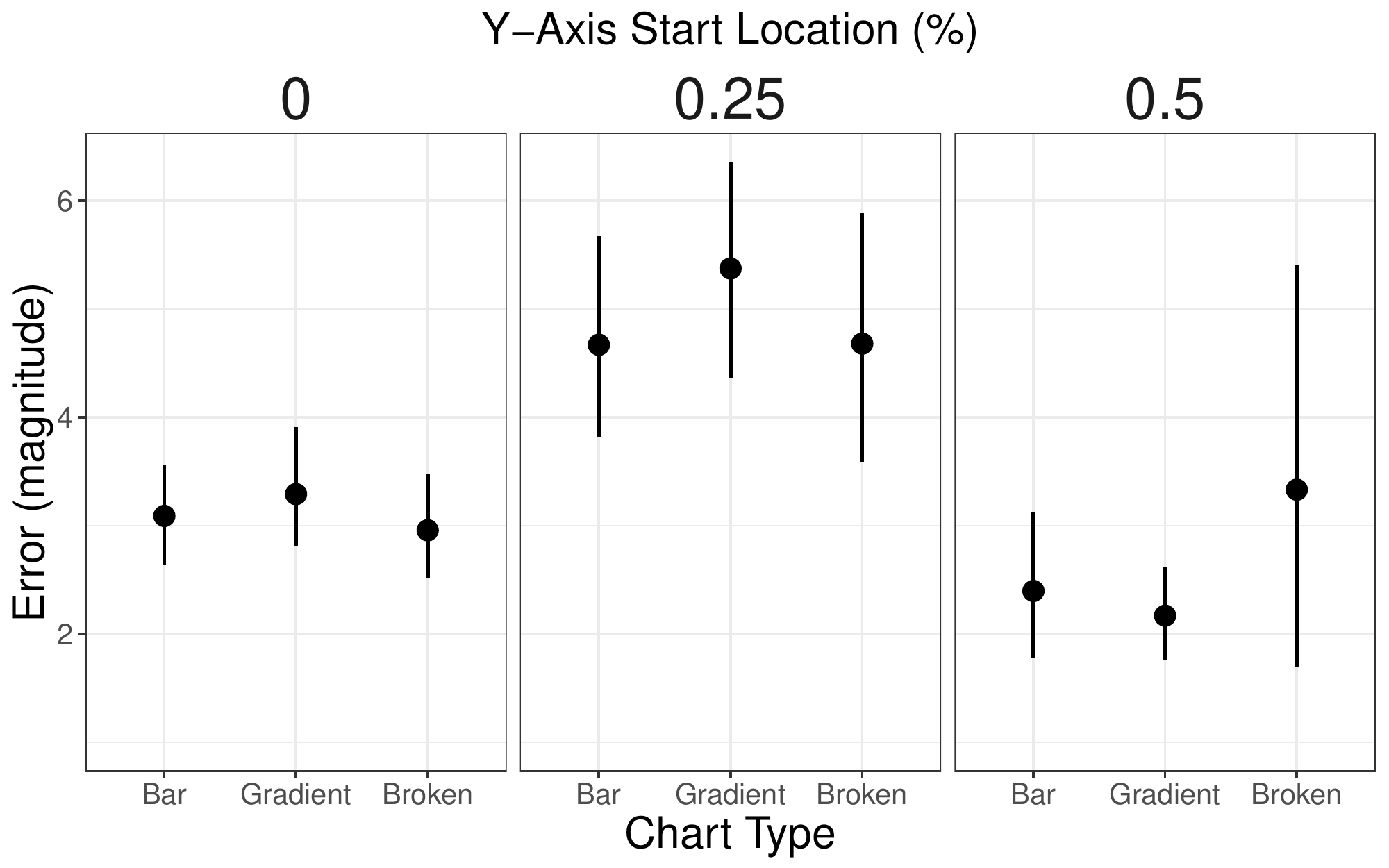}
    \label{fig:exp3avg}
    }
    \caption{Results from Experiments Three. Participants did not appear to reliably overestimate the trend in values across the chart, regardless of the level of truncation (\ref{fig:exp3trend}). However, there was more error in estimating individual values (\ref{fig:exp3avg}) when the axis began at $25\%$, which some participants reported as relating to the difficulty of transforming those valus back to the 0-100\% range. Error bars represent 95\% bootstrapped confidence intervals of the mean.}
    \label{fig:exp3errors}
\end{figure*}

The second experiment indicates that even designs that explicitly call attention to axis breaks do not have a noticeable impact on reducing the subjectively assessed trend. However, the relative unfamiliarity of the designs we chose, along with our decision to use crowd-sourced participants, suggests an alternative explanation: that our subject pool, regardless of design, simply ignores, discounts, or otherwise misreads the information on the y-axis. While we reiterate here that we do not believe that the effect of y-axis truncation is primarily a misreading of values (but is rather a visual exaggeration of an effect size), truncation has the side effect that a viewer inattentive to the axis labels would incorrectly decode the values in the chart. Even if the participants are \emph{aware} of the values, it is possible that by forcing participants to \emph{attend} more closely to specific numerical values would reduce the impact of the visual impact of the truncation as they consider the numerical difference between values.

In order to assess these possibilities, as well as disambiguate a \emph{bias} in trend from a \emph{misreading} of trend, we conducted a further experiment where we supplemented our existing qualitative rating task with a quantitative value task.

\subsubsection{Methods}
In this experiment we were interested in whether or not y-axis truncation results not just in qualitative increase in effect sizes but also misreading of values or if, conversely, attending to the specific value of numbers reduces this exaggeration. Therefore, we repeated the experimental design from Experiment Two, but included two additional tasks for each trial. Before the subjective rating task, participants were asked:
\begin{enumerate}
    \item ``What is the value of the first bar (0-100\%)''
    \item ``What is the value of the last bar (0-100\%)''
\end{enumerate}
Participants entered their guesses in text boxes. We did not directly ask for a numerical estimate of the slope to avoid an entangling between numerical guesses and the rating scale.

Using the answers to these estimation questions $Q_{first}$ and $Q_{last}$ compared to the actual values $X_{first}$ and $X_{last}$, we calculated two different error metrics roughly corresponding to errors in \emph{slope} or \emph{magnitude}. An error in slope would assert that truncation of the y-axis would cause participants to overestimate the slope of the resulting trend, and was calculated as the difference in estimated versus actual slope, $E_{slope} = |(Q_{last} - Q_{first}) - slope_{actual}|$. An error in magnitude would assume that the participants correct assessed the slope, but ignored the amount of truncation, inflating the size of the individual values, and was calculated as the average per-value estimation error, $E_{magnitude} = |(Q_{first} - X_{first})-(Q_{last} - X_{last} )| / 2$.

The experiment had the same factorial design as Experiment Two, for $3 \times 3 \times 2 \times 2 = 36$ $ + 8 $ validation $= 44$ total stimuli.

\subsubsection{Hypotheses}
We believed that explicitly soliciting value estimations would encourage participants to read the axis labels and therefore be more mindful of y-axis manipulations. If inattention to axis labels was a primary driver of the truncation bias, then our quantitative task, assuming good performance, would reduce it. Our second hypothesis was therefore that \textbf{severity would not differ across levels of truncation}; that is, the bias we observed would be reduced or eliminated in this version of the experiment.

Also under the assumption that participants would read the axis labels, and thus adjust their estimates accordingly, we assumed that \textbf{estimation error would also be similar across levels of truncation}, for both of the error metrics we calculated.

\subsubsection{Results}
We recruited 25 participants for this task (14 female, 11 male, $M_{age}=26.1$, $SD_{age}=9.2$). We paid participants \$4 for this task, for an effective hourly rate of $8.89$/hour. Participants had similar scores on the Galesic and Garcia-Retamero graphical literacy scale ($M_{correct}=11$, $SD_{correct}=1.6$), and a similar proportion answered the y-axis truncation item correctly ($21/25 = 84\%$). All participants met the inclusion criteria for accuracy on our engagement task, and we coded 56\% of participants as explicitly mentioning the y-axis manipulation.

We conducted an identical repeated measures ANOVA as the previous experiment, testing the effect of truncation level, visualization type, data size, and their interactions on perceived severity using the subset of trials where the y-axis truncation was $>0$. We also performed identically structured repeated measures ANOVA but where the response variable were our two value-estimation metrics $E_{slope}$ and $E_{magnitude}$

Our results fail to support our first hypothesis: as with the prior two experiments, \textbf{perceived severity was significantly different across levels of truncation} ($F(1,20) = 11$,$p=0.003$).  A post-hoc pairwise t-test with a Bonferroni correction found that all three levels of truncation had average perceived severity levels that were significantly different from each other; Fig. \ref{fig:exp2designs} shows this result, also revealing a similar pattern of responses between this experiment and experiment two.

Our results also fail to support our second hypotheses, with one exception. \textbf{Error in estimating trend was not significantly different across levels of truncation} ($F(1,20) = 0.002$, $p=0.96$). However, \textbf{Error in estimating individual values was significantly different across truncation levels} ($F(1,20) = 8.3$, $p=0.009$). A post-hoc pairwise t-test with a Bonferroni correction found that graphs with axes starting at $25\%$ had significantly higher error rates than other truncation levels, but no other differences. Participants specifically remarked upon the difficulty of this condition: P1:``I found out I had a much more difficult time interpreting charts when the scale was from 25 to 100, compared to any other scale'' and P9: ``Charts starting with 25\% are odds [\emph{sic}] in my opinion.'' This is potentially due to the relative difficulty in anchoring and converting values (for instance, half way up an axis that begins at $50\%$ is $75\%$, which can be used to anchor other estimates; halfway up a $25\%$ truncated axis is $62.5\%$, a far less convenient anchor point), but does not suggest that y-axis truncation creates a monotonic increase in error. 

\section{Discussion}
Our experimental results suggest that truncating the y-axis has a consistent and significant impact on the perceived importance of effect sizes. This qualitative bias occurs in both line charts and bar charts, as well as in bar charts that visually indicate either broken axes or the continuation of bars beyond the bounds of the chart. This bias is not merely a misreading of values, but seems to be connected to the visual magnification of differences.

These results suggest that, regardless of differences in the visual metaphors or encodings of line charts, there does not appear to be a significant \emph{practical} difference in the impact of truncation across different visualizations: the type of chart alone is not sufficient to shape guidelines around how to define charts. For the same data, the narrower the range of values in the y-axis, the larger the visual effect size and so the larger the subjective effect size. Different designs might provide more visual \emph{indications} that this exaggeration is occurring, but did not substantially alter the reported impact of the exaggeration.

Moreover, we cannot \emph{rely} on visual indicators of broken or truncated axes to counteract the exaggeration caused by y-axis truncation. Subjective judgments about effect size appear to be \emph{visual} rather than \emph{mathematical or statistical} judgments. Merely indicating that truncation has occurred, even in a prominent and unambiguous way, may not be sufficient to ``de-bias'' viewers of truncated charts. Surprisingly, The accurate estimation of values does not seem to counteract the visual magnification of difference.

However, we resist the interpretation of our experimental results to mean that, as Huff suggests~\cite{huff1993lie}, all charts with quantitative axes should include 0. The designer of the visualization, by selecting a y-axis starting point, has control over the subjective importance of the resulting differences~\cite{witt2019graph}. There is no \emph{a priori}, domain-agnostic \emph{ground truth} for how severe, important, or meaningful an effect size ought to be. We interpret our results as meaning that there is no obvious way for designers to \emph{relinquish the responsibility} of considering effect size in their charts. We reject the unequivocal dichotomy of ``honest'' and ``dishonest'' charts (for instance, as presented in Fig.~\ref{fig:badcharts}). 

\subsection{Limitations \& Future Work}
Our experiments focus on a limited set of designs to assess the impact of truncation on perceived effect size. We also focus on detecting the relative difference in subjective effect size across a few different levels of truncation, rather than attempting to fully model the complex relationship between slope, axis truncation, and perceived severity. It falls to future work to further explore the interplay of these variables.

Similarly, we tested only two potential designs for indicating axis truncation in bar charts as representatives of common classes of design interventions. Even of the designs we considered, we focused only on methods for static charts. Other methods using animation or interaction (such as in Ritchie et al.~\cite{ritchie2019}) could result in different patterns of subjective judgments by allowing the viewer to switch between truncated and non-truncated axes.

Wishing to avoid the complications involved in narrative or domain-focused crowdsourced studies (as discussed in Dimara et al.~\cite{dimara2018task}) our designs were presented in a relatively context-free manner. We believe that analysts in different domains have different internal models of effect size severity that would therefore not be captured in our results. We anticipate that different data domains and analytical contexts can impact the perceived importance or severity of effect sizes.

Connected with the issue of domain relevance is that of authority: visualizations from different sources or presented with different levels of perceived expertise or authority could produce differing patterns of judgment in different audiences. While there is ongoing work on understanding how visualizations persuade, and the rhetorical strategies that designers use to increase the persuasive power of visualization~\cite{pandey2014persuasive,kong2018frames}, a quantitative study of the persuasive power of y-axis truncation (especially for decision-making tasks) falls to future work.

\subsection{Conclusion}

Experts in information visualization and statistical graphics have produced conflicting advice on how harmful it is to start the y-axis of a chart from values other than 0. This conflict has often centered on the distinction between line graphs and bar charts, or on best practices for depicting axis breaks. Despite the claims that y-axis truncation is only ``deceptive'' for certain kinds of charts, or that explicit indication of axis breaks can ameliorate this ``deception,'' we find that the exaggeration introduced through truncation appears to persist across chart types and chart designs, and even when participants make accurate reports of the numbers they observe.

%Moreover, in many cases the ``deceptive'' practice of axis truncation is one of the most straightforward ways of communicating important and consequential effects and trends in the data. We therefore advise designers of visualizations to consider effect sizes in a domain- and audience-specific way, and choose chart designs that are effective for their communicative goals.

\section{Acknowledgments}
This work was supported by NSF awards CHS-1901485 and CHS-1900941. Thanks to Elsie Lee and Evan Anderson for assistance in qualitative coding.

% Balancing columns in a ref list is a bit of a pain because you
% either use a hack like flushend or balance, or manually insert
% a column break.  http://www.tex.ac.uk/cgi-bin/texfaq2html?label=balance
% multicols doesn't work because we're already in two-column mode,
% and flushend isn't awesome, so I choose balance.  See this
% for more info: http://cs.brown.edu/system/software/latex/doc/balance.pdf
%
% Note that in a perfect world balance wants to be in the first
% column of the last page.
%
% If balance doesn't work for you, you can remove that and
% hard-code a column break into the bbl file right before you
% submit:
%
% http://stackoverflow.com/questions/2149854/how-to-manually-equalize-columns-
% in-an-ieee-paper-if-using-bibtex
%
% Or, just remove \balance and give up on balancing the last page.
%
%\balance{}

% BALANCE COLUMNS
\balance{}

% REFERENCES FORMAT
% References must be the same font size as other body text.
\bibliographystyle{SIGCHI-Reference-Format}
\bibliography{template}

\end{document}

%\maketitle
%\tableofcontents
\begin{center}
    \huge{\plaintitle}\\
    \vspace{0.25 in}
    \large{Supplementary Material}    
        \vspace{0.5 in}
\end{center}

\normalsize

\section{Introduction}
This document is a supplement to the paper ``Truncating the Y-Axis: Threat or Menace?'' and includes additional analyses and figures excluded from the main paper for reasons of space. Consult our OSF project at \href{https://osf.io/gz98h/}{https://osf.io/gz98h/} for raw data tables, analysis scripts, and more reproducibility information.

\section{Full ANOVA Tables}

Our main experimental measure was a rating of the perceived severity of a trend in a chart purporting to show data over time. The main experimental manipulation was the starting location of the y-axis. See the main paper for additional details.

\subsection{Experiment One}
\begin{center}
    \begin{tabular}{|l|c|c|c|r|}
         \hline
         \multicolumn{5}{|c|}{Repeated Measures ANOVA Table (Perceived Severity)} \\
         \hline
         \textbf{Factor} & \textbf{df} & \textbf{dn} & \textbf{F} & \textbf{p} \\
         \hline
         \hline
         Truncation Level & 2 & 76 & 89 & 0.01E-20* \\
         \hline
         Visualization Type & 1 & 38 & 0.46 & 0.5 \\
         \hline
         Framing & 1 & 38 & 7.4 & 0.01* \\
         \hline
         Data Size & 1 & 38 & 5.5 & 0.02* \\
         \hline
         Truncation$\times$Vis & 2 & 76 & 0.67 & 0.52 \\
         \hline
         Truncation$\times$Framing & 2 & 76 & 0.59 & 0.55 \\
         \hline
         Vis$\times$Framing & 1 & 38 & 4.9 & 0.03* \\
         \hline
         Truncation$\times$Size & 2 & 76 & 0.24 & 0.79 \\
         \hline
         Vis$\times$Size & 1 & 38 & 1.4 & 0.24 \\
         \hline
         Framing$\times$Size & 1 & 38 & 2.0 & 0.16 \\
         \hline
         Trunc$\times$Vis$\times$Fram & 2 & 76 & 1.3 & 0.28 \\
         \hline
         Trunc$\times$Vis$\times$Size & 2 & 76 & 1.1 & 0.34 \\
         \hline
         Trunc$\times$Fram$\times$Size & 2 & 76 & 1.6 & 0.21 \\
         \hline
         Vis$\times$Fram$\times$Size & 1 & 38 & 2.3 & 0.14 \\
         \hline
         Trunc$\times$Vis$\times$Fram$\times$Size & 2 & 76 & 2.0 & 0.15 \\
         \hline
    \end{tabular}
\end{center}

\subsection{Experiment Two}
\begin{center}
    \begin{tabular}{|l|c|c|c|r|}
         \hline
         \multicolumn{5}{|c|}{Repeated Measures ANOVA Table (Perceived Severity)} \\
         \hline
         \textbf{Factor} & \textbf{df} & \textbf{dn} & \textbf{F} & \textbf{p} \\
         \hline
         \hline
         Truncation Level & 1 & 30 & 39 & 0.06E-7* \\
         \hline
         Visualization Type & 2 & 60 & 3.1 & 0.05 \\
         \hline
         Data Size & 1 & 30 & 3.1 & 0.09 \\
         \hline
         Truncation$\times$Vis & 2 & 60 & 2.5 & 0.09 \\
         \hline
         Truncation$\times$Size & 1 & 30 & 0.57 & 0.46 \\
         \hline
         Vis$\times$Size & 2 & 60 & 1.7 & 0.18 \\
         \hline
         Trunc$\times$Vis$\times$Size & 2 & 60 & 0.58 & 0.56 \\
         \hline
    \end{tabular}
\end{center}

\subsection{Experiment Three}

In Experiment Three, participants had to answer what the difference was between the last and the first data values. We calculated two forms of error: $E_{slope} = |(Q_{last} - Q_{first}) - slope_{actual}|$. and $E_{magnitude} = |(Q_{first} - X_{first})-(Q_{last} - X_{last} )| / 2$.

\begin{center}
    \begin{tabular}{|l|c|c|c|r|}
         \hline
         \multicolumn{5}{|c|}{Repeated Measures ANOVA Table (Slope Error)} \\
         \hline
         \textbf{Factor} & \textbf{df} & \textbf{dn} & \textbf{F} & \textbf{p} \\
         \hline
         \hline
         Truncation Level & 1 & 20 & 0.002 & 0.96 \\
         \hline
         Visualization Type & 2 & 40 & 0.80 & 0.45 \\
         \hline
         Data Size & 1 & 20 & 14 & 0.001* \\
         \hline
         Truncation$\times$Vis & 2 & 40 & 1.4 & 0.26 \\
         \hline
         Truncation$\times$Size & 1 & 20 & 0.67 & 0.42 \\
         \hline
         Vis$\times$Size & 2 & 40 & 1.6 & 0.21 \\
         \hline
         Trunc$\times$Vis$\times$Size & 2 & 40 & 0.95 & 0.40 \\
         \hline
    \end{tabular}
\end{center}

\begin{center}
    \begin{tabular}{|l|c|c|c|r|}
         \hline
         \multicolumn{5}{|c|}{Repeated Measures ANOVA Table (Magnitude Error)} \\
         \hline
         \textbf{Factor} & \textbf{df} & \textbf{dn} & \textbf{F} & \textbf{p} \\
         \hline
         \hline
         Truncation Level & 1 & 20 & 8.3 & 0.009* \\
         \hline
         Visualization Type & 2 & 40 & 0.17 & 0.85 \\
         \hline
         Data Size & 1 & 20 & 1.7 & 0.21 \\
         \hline
         Truncation$\times$Vis & 2 & 40 & 1.4 & 0.27 \\
         \hline
         Truncation$\times$Size & 1 & 20 & 0.25 & 0.63 \\
         \hline
         Vis$\times$Size & 2 & 40 & 0.05 & 0.95 \\
         \hline
         Trunc$\times$Vis$\times$Size & 2 & 40 & 0.29 & 0.75 \\
         \hline
    \end{tabular}
\end{center}

We also had the same rating task as the prior two experiments.

\begin{center}
    \begin{tabular}{|l|c|c|c|r|}
         \hline
         \multicolumn{5}{|c|}{Repeated Measures ANOVA Table (Perceived Severity)} \\
         \hline
         \textbf{Factor} & \textbf{df} & \textbf{dn} & \textbf{F} & \textbf{p} \\
         \hline
         \hline
         Truncation Level & 1 & 20 & 11 & 0.003* \\
         \hline
         Visualization Type & 2 & 40 & 0.22 & 0.80 \\
         \hline
         Data Size & 1 & 20 & 4.0 & 0.06 \\
         \hline
         Truncation$\times$Vis & 2 & 40 & 0.51 & 0.60 \\
         \hline
         Truncation$\times$Size & 1 & 20 & 0.11 & 0.75 \\
         \hline
         Vis$\times$Size & 2 & 40 & 1.2 & 0.31 \\
         \hline
         Trunc$\times$Vis$\times$Size & 2 & 40 & 0.52 & 0.59 \\
         \hline
    \end{tabular}
\end{center}

\section{Additional Figures \& Analyses}

\subsection{Raw Responses}

The main task in all three experiments was answering a question about the perceived severity of a trend in a data, given the chart used to show the data (and the starting point of the y-axis of that chart). This question was answered as a 5-point rating scale. While we mainly report on the central tendencies of this rating scale in the main paper, we recognize that rating scales are often better conceptualized as ordinal rather than purely numerical data. We include the pattern of raw responses here.

    \begin{figure*}
    \centering
    \subfloat[Experiment One]{
    \includegraphics[width=0.24\textwidth]{exp1Stacked.pdf}
    \label{fig:exp1stack}
    }
    ~
    \subfloat[Experiment Two]{
    \includegraphics[width=0.3\textwidth]{exp2Stacked.pdf}
    \label{fig:exp2stack}
    }
    ~
    \subfloat[Experiment Three]{
    \includegraphics[width=0.3\textwidth]{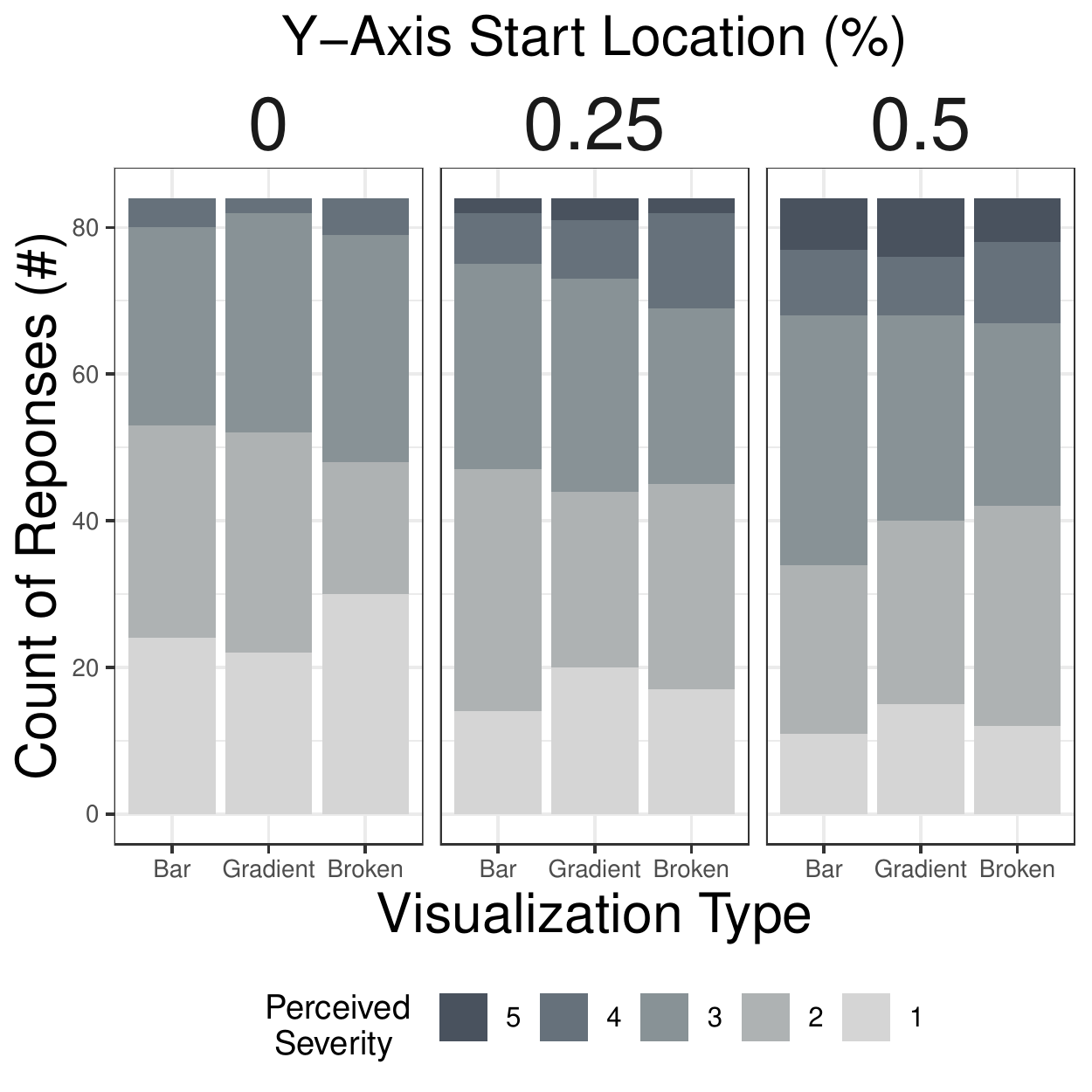}
    \label{fig:exp3stack}
    }
    \caption{Stacked bar charts of all raw responses from all three experiments. The darker the value, the more severe the effect was judged as being. Note that the actual set effect sizes were numerically identical across all conditions, so if the starting location of the y-axis had no impact, each bar would have a similar proportion of responses.}
    \label{fig:stacks}
\end{figure*}

\subsection{Response Time}

We measured response time from when the participant clicked the ``Ready'' button, to when they finalized their choice with the ``Confirm'' button. Error bars are 95\% bias-corrected bootstrapped confidence intervals. Note that we did not instruct participants to answer as quickly as possible (although on crowdworking platforms there is an inherent pressure to complete tasks as quickly as possible), nor did we alert them that we were recording this timing information, or forbid participants from pausing the task to complete other activities. As such, the values are highly variable.

\begin{figure*}
    \centering
    \subfloat[Experiment One]{
    \includegraphics[width=0.4\textwidth]{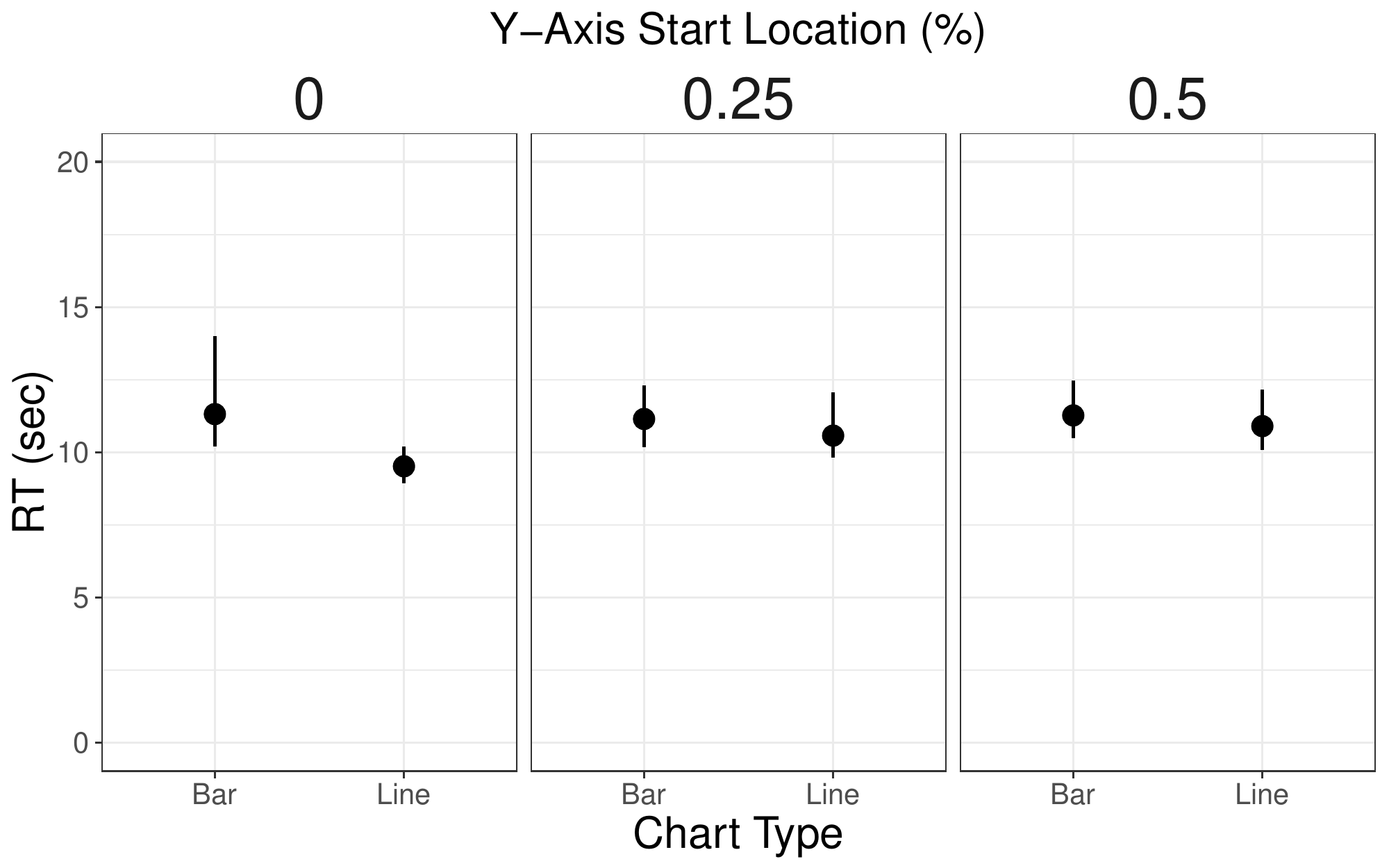}
    \label{fig:exp1rt}
    }
    ~
    \subfloat[Experiment Two]{
    \includegraphics[width=0.4\textwidth]{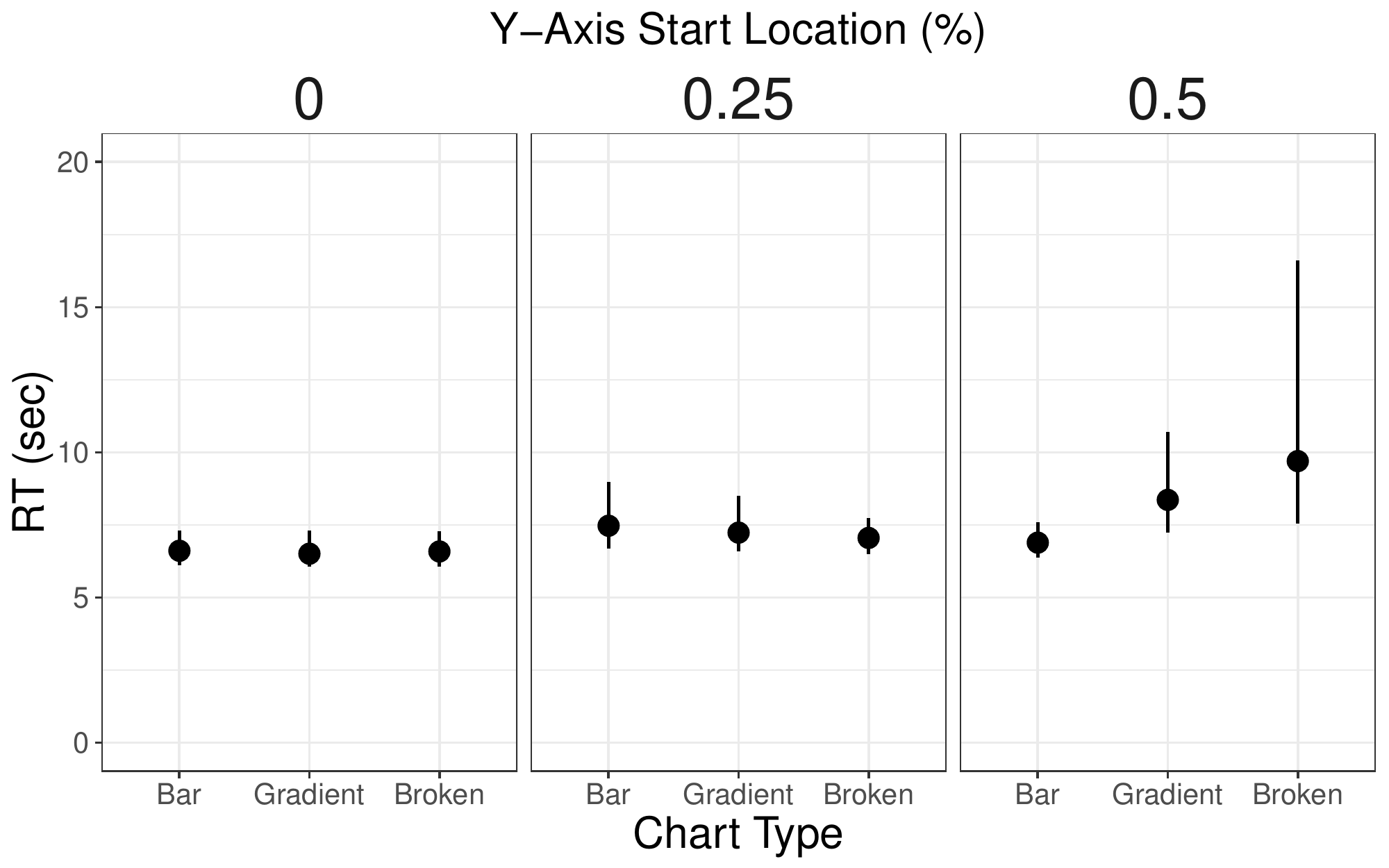}
    \label{fig:exp2rt}
    }
    \caption{Response time for Experiments One and Two. Error bars are 95\% bias-corrected bootstrapped confidence intervals.}
    \label{fig:rt}
\end{figure*}

\begin{figure}
    \centering
    \includegraphics[width=0.4\textwidth]{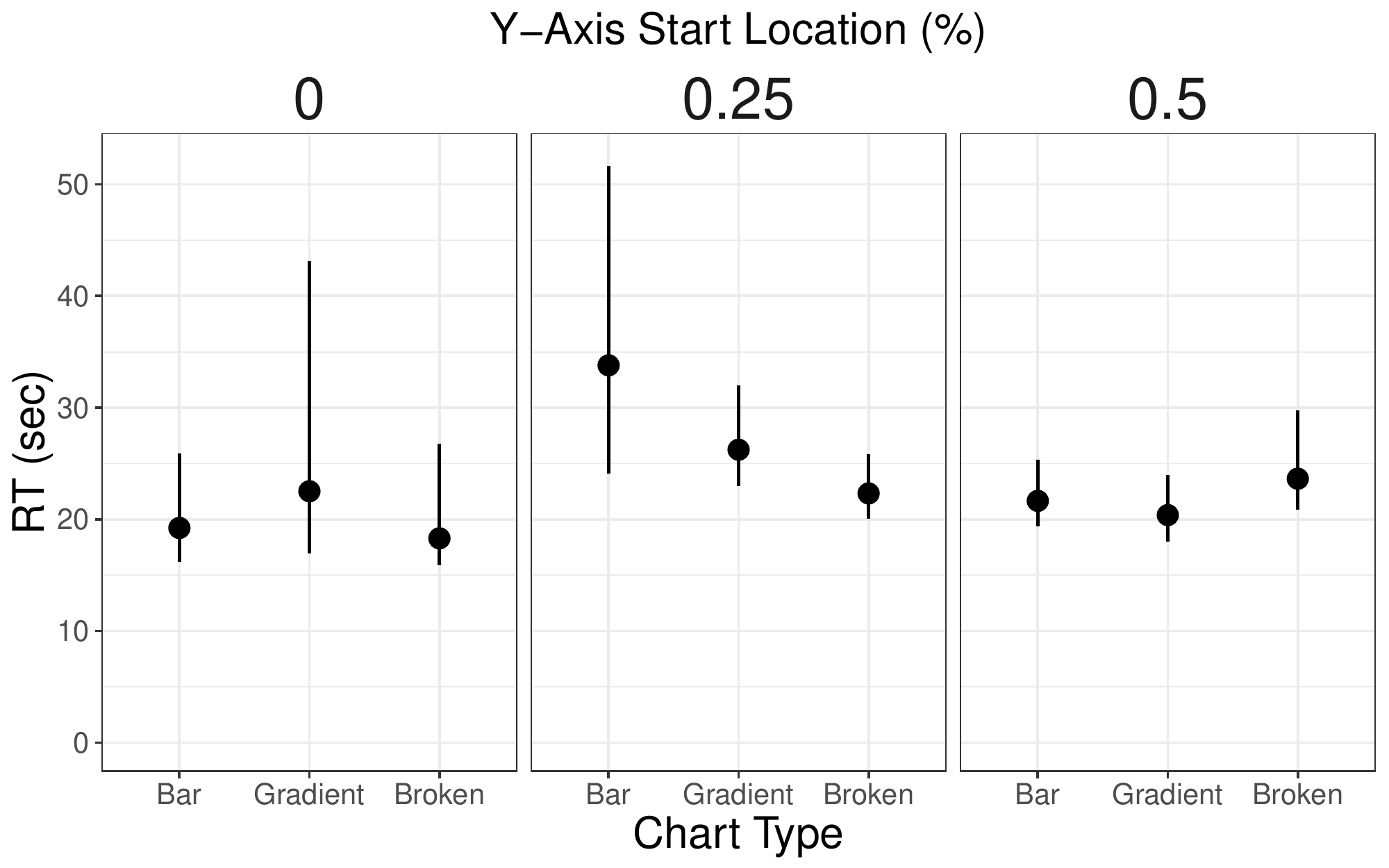}
    \label{fig:exp3stack}
    \caption{Response time for Experiment Three. Note the differing y-axis between this chart and \protect\autoref{fig:rt}: in Experiment Three there was a third, much more involved numerical task, so RTs are not comparable here to the other experiments.}
    \label{fig:stacks3}
\end{figure}

\begin{center}
    \begin{tabular}{|l|c|c|c|c|}
         \hline
         \multicolumn{5}{|c|}{Repeated Measures ANOVA Table} \\
         \hline
         \textbf{Factor} & \textbf{df} & \textbf{dn} & \textbf{F} & \textbf{p} \\
         \hline
         \hline
         Truncation Level & 2 & 76 & 1.4 & 0.25 \\
         \hline
         Visualization Type & 1 & 38 & 4.0 & 0.05 \\
         \hline
         Framing & 1 & 38 & 17 & 0.0002* \\
         \hline
         Data Size & 1 & 38 & 3.5 & 0.07 \\
         \hline
         Truncation$\times$Vis & 2 & 76 & 1.15 & 0.32 \\
         \hline
         Truncation$\times$Framing & 2 & 76 & 0.46 & 0.62 \\
         \hline
         Vis$\times$Framing & 1 & 38 & 0.003 & 0.96 \\
         \hline
         Truncation$\times$Size & 2 & 76 & 0.02 & 0.97 \\
         \hline
         Vis$\times$Size & 1 & 38 & 0.60 & 0.44 \\
         \hline
         Framing$\times$Size & 1 & 38 & 1.1 & 0.30 \\
         \hline
         Trunc$\times$Vis$\times$Fram & 2 & 76 & 0.53 & 0.59 \\
         \hline
         Trunc$\times$Vis$\times$Size & 2 & 76 & 0.76 & 0.47 \\
         \hline
         Trunc$\times$Fram$\times$Size & 2 & 76 & 1.2 & 0.32 \\
         \hline
         Vis$\times$Fram$\times$Size & 1 & 38 & 0.18 & 0.67 \\
         \hline
         Trunc$\times$Vis$\times$Fram$\times$Size & 2 & 76 & 1.1 & 0.33 \\
         \hline
    \end{tabular}
\end{center}

\subsection{Strategy Information}

After the main task, participants were asked to report on their strategy, and if they noticed anything about the graphs they were looking at. We were specifically looking for if the participants reported noticing that the y-axes of the charts had different start points. For each experiment, two coders independently gave a binary code to the free text responses if they judged that the participants' responses gave such an indication. The coders then met to assess and rectify mismatches. In this section we report on the impact of these codes on the size of the impact of y-axis truncation on subjective impressions of trend.

\begin{figure}
    \centering
    \includegraphics[width=0.6\textwidth]{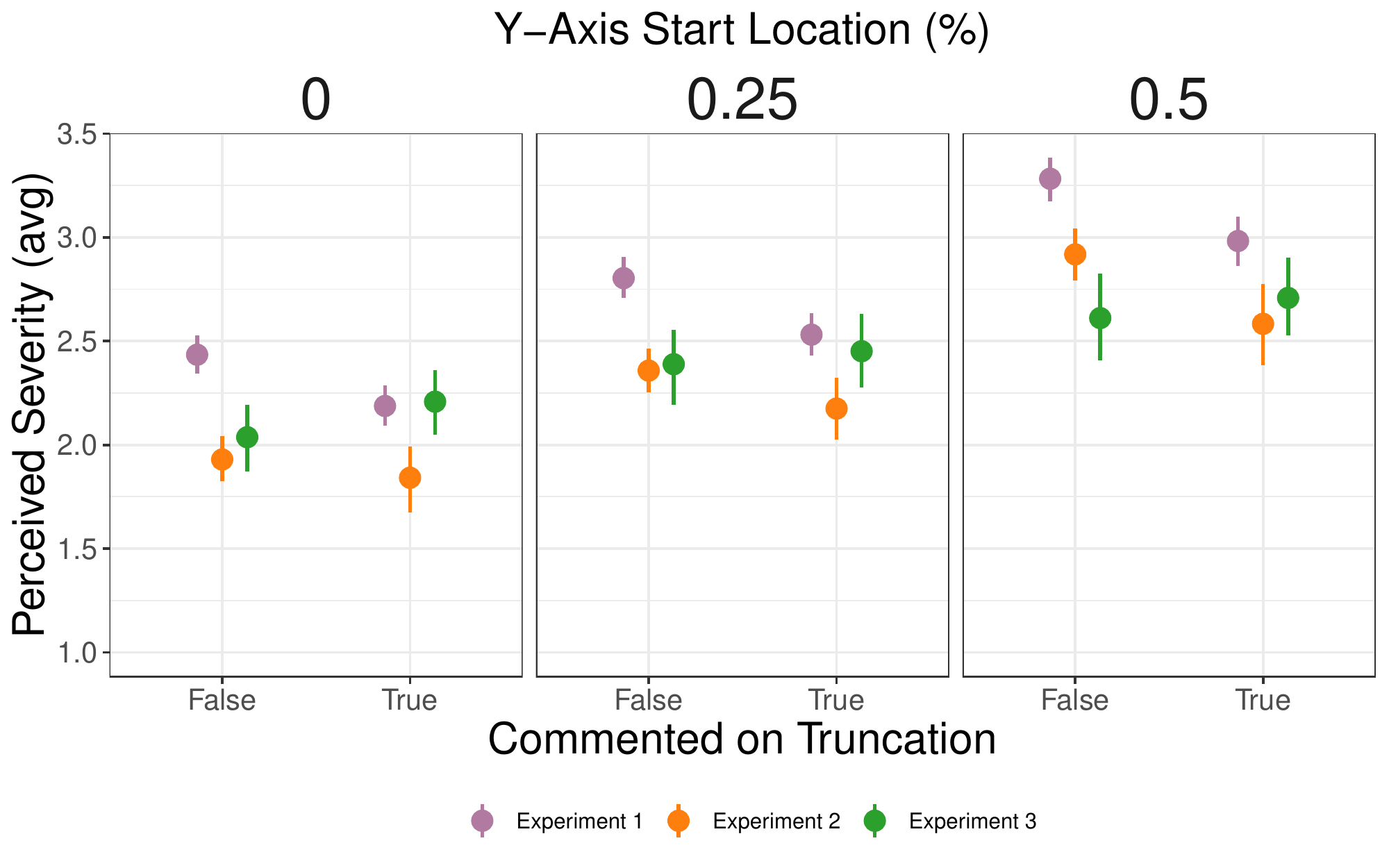}
    \label{fig:exp3stack}
    \caption{Impact of explicit mention of the y-axis manipulation on perceived severity of trend. In all cases, across all experiments there was a consistent impact of the y-axis start location, even for ``saavy'' participants. In Experiments One and Two there was a slight decrease in perceived severity for those who commented on y-axis truncation. But there was a slight increase in Experiment Three. (Although note that in Experiment Three participants had to answer numerical questions about the data, so it's possible that the truncation was so ``obvious,'' as it was central to the task, that it was not diagnostic to mine responses for its presence or absence. Error bars are 95\% bias-corrected bootstrapped confidence intervals.}
    \label{fig:stacks3}
\end{figure}

\pagebreak
\vfill
\footnotesize        

%\bibliographystyle{SIGCHI-Reference-Format}
%\bibliography{template}